\documentclass[a4paper,11pt]{article}
\pdfoutput=1 
\usepackage{jcappub} 
\usepackage[T1]{fontenc} 
\usepackage{multirow}
\usepackage{longtable}
\usepackage{tabularx}
\usepackage{booktabs}
\graphicspath{{plots/}}                      
\usepackage{ifthen}
\usepackage{siunitx}
\usepackage{xcolor}
\usepackage{hyperref}
\usepackage{url}
\usepackage{mathrsfs}
\usepackage{amsmath, amssymb}
\usepackage{caption}
\usepackage{subcaption} 
\usepackage{titlesec}
\usepackage{array}
\usepackage{float}
\usepackage{afterpage}
\usepackage[normalem]{ulem}
\newcolumntype{?}[1]{!{\vrule width #1}}

\begin{document}
\title{\boldmath Particle production during inflation: constraints expected from 
redshifted 21 cm observations from the epoch of reionization} 

\author[1]{Suvedha Suresh Naik,}
\author[2,3]{Pravabati Chingangbam,}
\author[1]{Kazuyuki Furuuchi}

\affiliation[1]{Manipal Centre for Natural Sciences,
Centre of Excellence, Manipal Academy of Higher Education,
Dr. T.M.A. Pai Planetarium Building, Manipal 576 104, Karnataka, India}
\affiliation[2]{Indian Institute of Astrophysics,
	Koramangala II Block, Bangalore 560 034, India}
\affiliation[3]{School of Physics, Korea Institute for Advanced Study, 85 Hoegiro, Dongdaemun-gu, Seoul, 02455, Korea}

\emailAdd{suvedha.nail@learner.manipal.edu}
\emailAdd{prava@iiap.res.in}
\emailAdd{kazuyuki.furuuchi@manipal.edu}

\abstract{
We examine a type of feature in the primordial scalar power spectrum, namely, 
the bump-like feature(s) that arise as a result of burst(s) of particle production during inflation.  
The latest CMB observations by {\it Planck} 2018 
can accommodate the imprints of 
such primordial features. 
In the near future, observations of redshifted 21~cm signal {from the Epoch of Reionization (EoR)} can
put additional constraints on  
inflation models
{by exploiting} the expected tomographic information 
across a wide range of co-moving wave-numbers. 
In this work, we study the potential of upcoming 
observational data from SKA-Low to 
constrain the parameters of the primordial power spectrum with bump-like features. 
{We use simulated mock data 
expected from SKA-Low,}
with uncertainties estimated from different foreground removal models,
and 
{constrain}  the parameters of primordial features
within a Bayesian 
framework. We study two scenarios:
{in the first scenario,  
where the astrophysical parameters 
relevant to the EoR 
are known,} 
we find that 21~cm power spectra  
do have the potential to probe the 
primordial bump-like features. 
{As the input amplitude of the bump is decreased  below roughly 10\% of the amplitude of the primordial power spectrum without the bump, the uncertainties in the recovered values for both amplitude and location of the bump are found to increase, 
and the recovered values of the location of the bump also get increasingly more biased towards higher values. 
Hence, it becomes harder to constrain these parameters.} 
{In the second scenario}, we analyze the achievable constraints on primordial features when 
two of the EoR parameters, namely, minimum halo mass and ionizing efficiency, are uncertain. 
We find that the effect of the bump on the profile and the amplitude of the 21 cm power spectrum is distinct from the impact of changing the astrophysical parameters, and hence they may potentially be distinguished. }
\maketitle
\section{Introduction}
\label{sec:intro}
Inflation 
\cite{Guth:1981,LINDE:1982,Albrecht:1982,STAROBINSKY198099,stato:1981,kazanas:1980} 
is regarded as a standard 
framework for constructing theoretical models that  
can explain 
the fine-tuned initial conditions
of 
the Big Bang model. 
The quantum fluctuations generated during inflation 
are considered as the origin of primordial
density perturbations. 
These density fluctuations 
evolve in time 
and result in the structures that we observe in cosmological data. 
The observational data is then expected to contain imprints of the physics that operated during the primordial epoch.

During  the last few decades, substantial effort has been made 
to probe primordial density perturbations
using the Cosmic Microwave Background (CMB)
anisotropies and 
{fluctuations in the spatial distribution of galaxies.}
The 
{current status is that} the observations are consistent with the concordance
$\Lambda$CDM model. In the concordance model, 
the primordial density fluctuations are described by 
a nearly scale-invariant primordial power spectrum. The CMB measurements from \textit{Planck} \cite{Akrami:2018vks,Akrami:2018odb}
have tightly constrained the parameters of
the concordance $\Lambda$CDM model. 
{However, the data contains larger uncertainties towards the large and small scales 
and can accommodate primordial power spectra that exhibit deviations from the nearly scale-invariant form.  
On the theoretical modelling side, the inflationary framework
accommodates a variety of theoretical models
which can be investigated by probing their unique signatures
or ``features'' that manifest themselves as deviations from near scale-invariance 
of the primordial power spectrum at different scales. 
{Probing the primordial features}
using 
observations across a wide range of 
wave-numbers
will {further tighten the constraints on inflationary models.}}

A class of models involving bursts of
particle production during inflation predicts bump-like 
features on the primordial power spectrum
\cite{Chung:1999ve,Barnaby:2009mc,Barnaby:2009dd,Pearce:2017bdc}.
In inflation models based on higher dimensional gauge theories,
such bursts of particle production may occur naturally
\cite{Furuuchi:2015foh,Furuuchi:2020klq,Furuuchi:2020ery}
motivating the search for the signatures of 
bump-like features in cosmological observations.
Recently, the presence of such features was 
investigated with the latest CMB data from 
the \textit{Planck} satellite \cite{Naik:2022mxn}.
The \textit{Planck} data puts constraints on the amplitude of the 
bump-like features on co-moving wave-numbers 
$0.0002 \lesssim k({\rm Mpc}^{-1}) \lesssim 0.15$, 
which in turn constrains  
the coupling parameter responsible for 
particle production during inflation.
The bump-like features were also investigated with 
the galaxy two-point correlation functions \cite{Ballardini:2022wzu},
future CMB observations and galaxy surveys \cite{Chandra:2022utq}.

Neutral hydrogen, being the most abundant baryonic matter, 
has the potential to probe the unexplored windows of the universe
via the 21~cm signal from its hyperfine transition
\cite{Furlanetto:2006jb,Furlanetto:2009astro2010S..82F,Pritchard:2012RPPh...75h6901P}.  
The 21~cm fluctuations during the cosmic dawn are 
direct tracers of the underlying matter distribution and
hence ideal for probing the 
primordial density fluctuations \cite{Loeb:2003ya}.
The upcoming redshifted 21~cm observations from 
instruments such as 
the Hydrogen Epoch of Reionization Array 
(HERA) \cite{hera}
and Square Kilometre Array 
(SKA) \cite{ska}
are expected to significantly improve 
our understanding of physics at different epochs of the universe.  
Extracting the signatures of primordial features 
from the future redshifted 21~cm signal
will be challenging due to the uncertainty
in astrophysical processes, 
{foreground contamination and  
instrumental noise. }
However, the tomographic
study of redshifted 21~cm signal
provides the hope to probe the early universe
when information from different statistical measures such as
sky-averaged 21~cm signal, power spectrum and 
higher order correlators are combined. 
{Previous studies have investigated prospects of probing the primordial
universe with near-future redshifted 21~cm observations. 
In particular, }the potential of the highly redshifted 21~cm signals to constrain 
the primordial oscillatory features was studied in \cite{Chen:2016zuu}.
The primordial features of resonant and step inflation
models were investigated in \cite{Xu:2016kwz} with the expected 21~cm intensity mapping 
observations by SKA1-Mid\footnote{The SKA’s mid-frequency instrument
covering 350 MHz to 14 GHz \cite{SKA:2018ckk}.} and 
Tianlai observations \cite{tianlai}.
The primordial features of kink, 
step and warp {types} were studied in \cite{Ballardini:2017qwq}
with intensity mapping from SKA1-Mid and 
the expected photometric surveys from LSST \cite{lsst}.
However, the studies \cite{Chen:2016zuu,Xu:2016kwz,Ballardini:2017qwq} have 
used the Fisher information matrix to forecast the uncertainties
of their model parameters.
The Fisher matrix approach assumes ideal observations
with the uncertainty containing the description of all sources of errors. 
Recently, the position of the absorption trough 
in the global 21~cm signal from the EDGES experiment \cite{Bowman:2018yin}
was used as a probe of
the amplitude of the small-scale primordial power spectrum
\cite{Yoshiura:2018zts,Yoshiura:2019zxq}.
To 
{the best of} 
our knowledge, a detailed analysis of the primordial
features with more realistic observations from future
21~cm telescopes has not been carried out.


In this paper, 
we perform Bayesian analysis against
the simulated data expected from  
near-future redshifted 21 cm observations
to probe 
particle productions during inflation.
{As the redshifted 21~cm observations are expected to probe
the matter distribution across various co-moving
wave-numbers, 21~cm surveys 
will provide information in addition to what is available from CMB and LSS.
In particular, 
we study the signatures of bump-like features in the 
redshifted 21~cm signal from the Epoch of Reionization (EoR),
targeted in SKA-Low\footnote{The SKA’s low-frequency instrument
covering 50 to 350 MHz \cite{SKA:2018ckk}.}.} 
We generate the uncertainty expected from SKA-Low
by considering their antenna coordinates 
to mimic more realistic observations
than 
{what is} obtained from Fisher matrix formalism.
As the cosmic 21~cm signal is sensitive to initial conditions
and the underlying astrophysical parameters at different redshifts, 
one needs a systematic exploration of these parameters. 
To obtain detailed information on the constrained parameters
and possible degeneracies, 
we use the Markov Chain Monte Carlo (MCMC) approach
to numerically evaluate the posterior probability distributions 
of the parameters.
We examine the ability of SKA-Low
to recover the parameters of primordial features
within a Bayesian MCMC framework. 
We investigate the possibility of the existence of primordial features
in the co-moving wave-number {range} 
$0.1 \lesssim k({\rm Mpc}^{-1}) \lesssim 1.0$. 
We find that SKA-Low {does have} the potential 
to probe the presence of primordial features
when the astrophysical parameters are fixed
to their benchmark values considered in this work. 
{We also discuss} the challenges in recovering the parameters of primordial
features in the case of uncertain astrophysical parameters.

This article is organized as follows: 
the primordial power spectrum motivated 
by particle production during inflation
is briefly described in section~\ref{sec:model}. 
In section~\ref{sec:method}, we provide the 
methodology followed in this work. 
Mainly, this section describes the simulation of redshifted 21~cm signals, 
the generation of noise power spectrum from the SKA-Low
and the details of MCMC sampling. 
In section~\ref{sec:results}, we present 
our results obtained from the Bayesian analysis. 
We summarize the study and discuss future directions 
in section~\ref{sec:Discussion}.

\section{Primordial features due to particle productions 
during inflation}
\label{sec:model}
{\em The concordance model}:
Before discussing the primordial features investigated in this work, 
we provide the primordial power spectrum from the 
concordance $\Lambda$CDM model,
which will be compared with the primordial 
power spectrum with features. 

The concordance $\Lambda$CDM model assumes a scalar power spectrum parameterized by
the amplitude of the scalar perturbations $A_s$
and the scalar spectral index $n_s$:
\begin{equation}\label{eq:PS_PowerLaw}
P_s(k)
=
A_s \left(\frac{k}{k_0}\right)^{n_s-1}\,,
\end{equation}
where $k$ is the co-moving wave-number and the pivot scale $k_0(\text{Mpc}^{-1})$
is
{chosen to be} $0.05$ ~\cite{Akrami:2018vks}.
%

\noindent{\em Models with primordial features due to particle production during inflation:} 
We study a class of inflation models 
\cite{Chung:1999ve,Barnaby:2009mc,Barnaby:2009dd,Pearce:2017bdc} in which 
the inflaton field $\phi$ is coupled to a real 
scalar field $\chi$ through the interaction term 
\begin{equation}
g^2
(\phi - \phi_0)^2 \chi^2 \,,
\label{eq:g_square}
\end{equation}
where $g$ is the dimensionless coupling constant.
When the inflaton field value crosses $\phi = \phi_0$, 
a burst of $\chi$ particle production occurs as they
become instantaneously massless. 
{When such an event occurs} 
during the observable range of $e$-folds of inflation, 
{it} 
appears
in the primordial power spectrum 
as a 
bump-like feature. 
Inflation models based on gauge theory in
higher dimensions naturally 
give rise to the coupling of the form eq.~\eqref{eq:g_square}
\cite{Furuuchi:2015foh,Furuuchi:2020klq,Furuuchi:2020ery}.
The coupling parameter $g$ can be constrained by studying 
the imprints of bump-like features on cosmological observations. 

In \cite{Pearce:2017bdc}, the dominant and subdominant
contributions to the power spectrum were
calculated analytically with one-loop approximations.
The bounds on the coupling constant $g$ 
as studied in \cite{Pearce:2017bdc} are mentioned below. 
The condition that the two-loop corrections are subdominant
with respect to the leading one-loop contribution
requires 
\begin{equation}
    g^2 \lesssim 3\,.
    \label{eq:g_upper}
\end{equation}
The condition that the mass of $\chi$ evolves quickly 
enough during the event of particle production requires 
\begin{equation}
    g^2 \gg 10^{-7}\,.
    \label{eq:g_lower}
\end{equation}

The primordial bump-like features 
produced due to  particle production during inflation
were fitted using numerical results in 
\cite{Barnaby:2009dd,Barnaby:2009mc}.
In this work, we parameterize the 
primordial power spectrum by considering the 
latest analytical results \cite{Pearce:2017bdc}
and given by
\begin{equation}
P_s(k)
=
A_s \left(\frac{k}{k_0}\right)^{n_s-1}
+
A_{\rm I} \sum_i \left(\frac{f_1(x_i)}{f_1^{\rm max}}\right)
+
A_{\rm II} \sum_i \left(\frac{f_2(x_i)}{f_2^{\rm max}}\right)\,.
\label{eq:PS_bump}
\end{equation}
%
For an inflation model involving a coupling to a real scalar field, 
as in eq.~\eqref{eq:g_square},
the amplitudes depend on the model parameter $g$ as\footnote{The 
dependence of $A_{\rm I}$ and $A_{\rm II}$ on $g$ is 
the case of a real scalar field.
A factor of two should be multiplied appropriately for the case of a complex scalar field.
{
For models based on gauge theory in higher dimensions
\cite{Furuuchi:2015foh,Furuuchi:2020klq,Furuuchi:2020ery},
the interaction of the form eq.~\eqref{eq:g_square}
arises from the minimal coupling of a gauge field 
with a charged scalar field that is complex, 
with $g$ being the 4D gauge coupling.}}
\begin{align}
A_{\rm I} &\simeq 6.6\times 10^{-7} g^{7/2} \,,\label{eq:A1}\\
A_{\rm II} &\simeq 1.1 \times 10^{-10} g^{5/2} \ln\left(\frac{g}{0.0003}\right)^2\,. \label{eq:A2}
\end{align}
%
%
\begin{figure}[tbp]
	\centering 
        \begin{subfigure}{.45\textwidth}
            \includegraphics[width=0.9\textwidth]{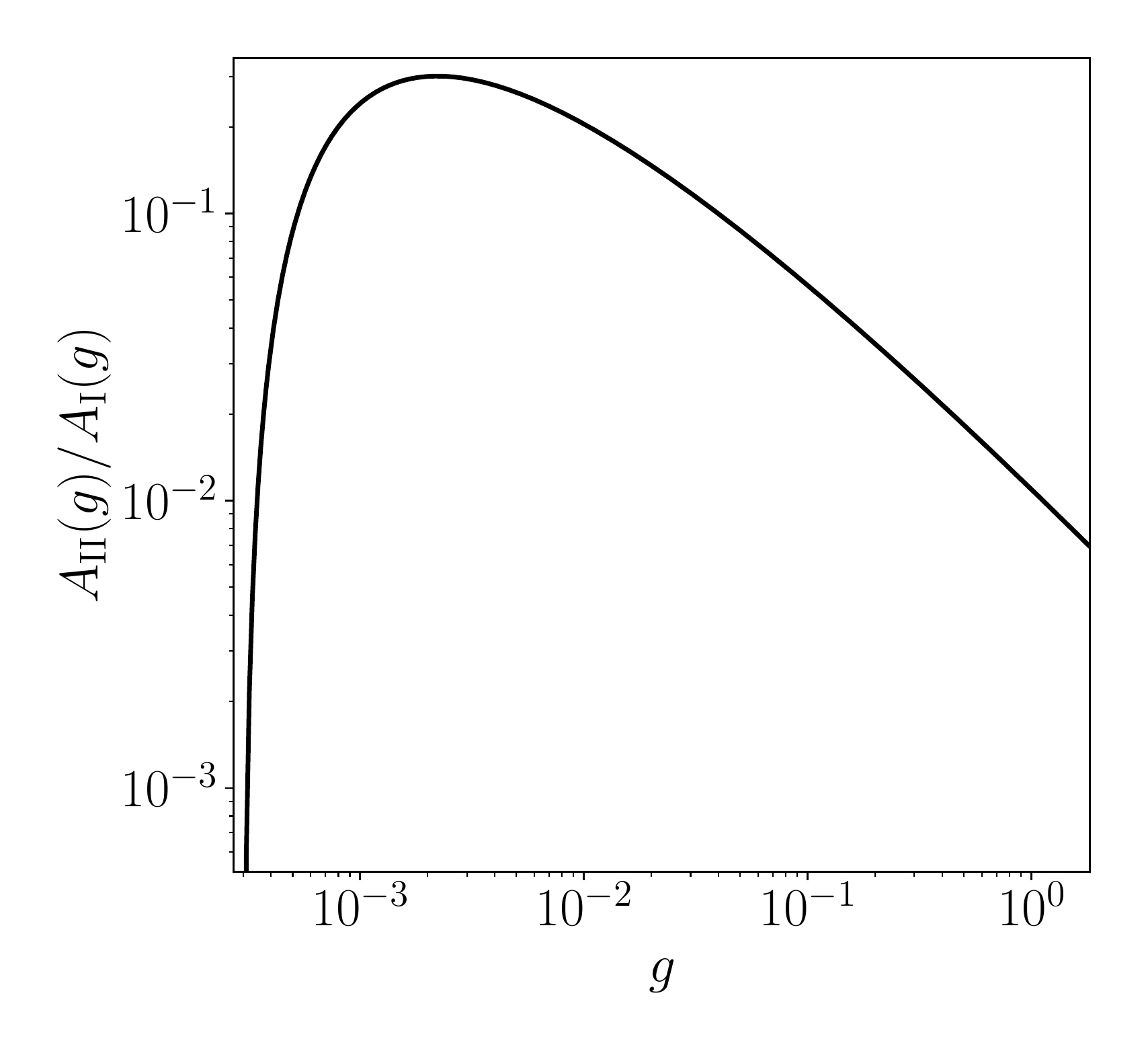}
            \caption{}
            \label{fig:A2A1}
        \end{subfigure}
	\begin{subfigure}{.45\textwidth}
            \includegraphics[width=0.9\textwidth]{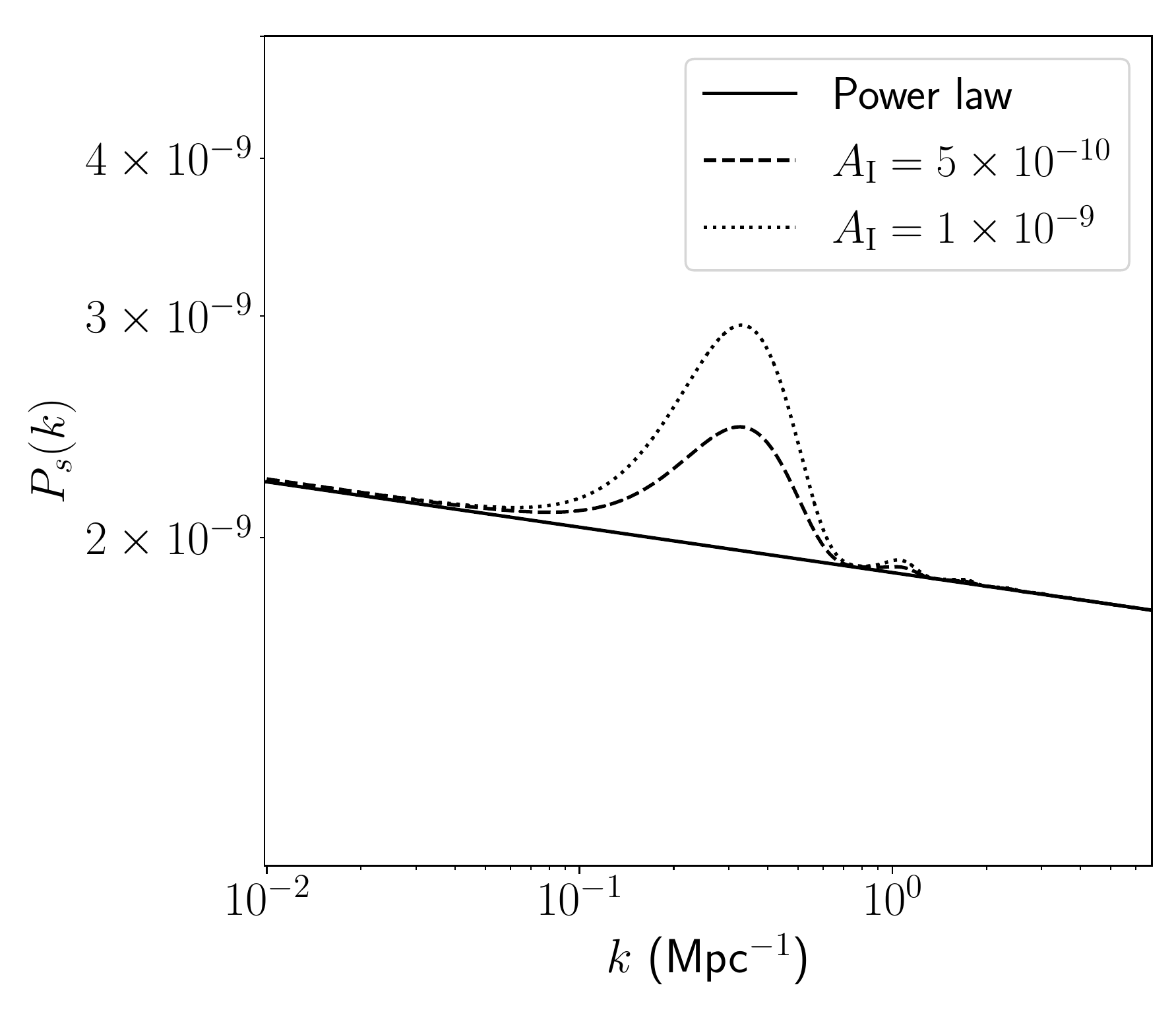}
            \caption{}
            \label{fig:singlebump}
        \end{subfigure}
	\caption{(a) The ratio of the amplitudes of the 
        subdominant feature $A_{\rm II}$ (eq.~\eqref{eq:A2}) to the dominant feature 
        $A_{\rm I}$ (eq.~\eqref{eq:A1}) as a function of $g$ within the bounds given by 
        the conditions \eqref{eq:g_upper} and \eqref{eq:g_lower}.
        (b) Single bump models with $k_b = 0.1({\rm Mpc}^{-1})$ 
	(the peaks occur at $3.35\times k_b$), 
	and $A_{\rm I}$ set at $1.0\times10^{-9}$ ({\it dotted}) and
	$5.0\times10^{-10}$ ({\it dashed}).
	The primordial power spectrum for the power law is plotted
	in a solid line.
        }
\end{figure}
In figure~\ref{fig:A2A1}, we plot the ratio of equations \eqref{eq:A1} and \eqref{eq:A2}
as a function of the parameter $g$ within the bounds given by \eqref{eq:g_upper} and \eqref{eq:g_lower}. 
The figure indicates that $A_{\rm II}$ is subdominant compared to $A_{\rm I}$
in the given range of $g$. 

The scale dependence of the dominant and subdominant contributions are given by the 
dimensionless functions
\begin{align}
f_1(x_i) &\equiv
\frac{\left[\sin(x_i)-{\rm SinIntegral}(x_i)\right]^2}{x_i^3} \,,\label{eq:f1}\\
f_2(x_i) &\equiv
\frac{-2x_i\cos(2x_i)+(1-x_i^2)\sin(2x_i)}{x_i^3}\,, \label{eq:f2}
\end{align}
where $x_i \equiv \frac{k}{k_i}$,
and $\text{SinIntegral}(x)=\int_0^{x}\frac{\sin z}{z}dz$.
The peaks of the functions $f_1(x)$ and $f_2(x)$
evaluate to $f_1^{\rm max} \simeq 0.11$ and $f_2^{\rm max}\simeq0.85$, respectively.
The parameter
$k_i({\rm Mpc}^{-1})$
is related to the location of the $i^{\rm th}$ feature on the primordial
power spectrum.
The peak of the $i^{\rm th}$ dominant function 
occurs at 
\begin{equation}
    k_{p,i} \simeq 3.35\times  k_i\,,
    \label{eq:kpeak}
\end{equation}
and the subdominant function 
occurs at $x_i \simeq 1.25$. 

In this work, we focus on a simple scenario of a single burst of 
particle production during inflation.
The primordial power spectrum for a
single bump having an amplitude $A_{\rm I}$ 
located at the scale $k_b$ (i.e. the peak at $\sim 3.35\times k_b(\text{Mpc}^{-1})$)
is given by
\begin{equation}\label{eq:single_bump}
P_s(k)
=
A_s \left(\frac{k}{k_0}\right)^{n_s-1}
+
A_{\rm I} \left(\frac{f_1(k/k_b)}{f_1^{\rm max}}\right)
+
A_{\rm II} \left(\frac{f_2(k/k_b)}{f_2^{\rm max}}\right)\,.
\end{equation}
The primordial power spectrum for a single bump model
is shown in figure~\ref{fig:singlebump}.
%

\section{Methodology}
\label{sec:method}
This section describes our method for
investigating the primordial bump-like features
with the redshifted 21~cm observations expected from SKA-Low. 

\subsection{Redshifted 21~cm line as a probe of the early universe}
\label{subsec:21cm}
The redshifted $21$ cm line can be a valuable probe of the high redshift universe, 
including the dark ages and the EoR 
\cite{Furlanetto:2006jb,Furlanetto:2009astro2010S..82F,Pritchard:2012RPPh...75h6901P,Loeb:2003ya,Mesinger:2019cosm.book}.
The observable quantity is the differential brightness temperature at redshift $z$, which quantifies
the change in brightness temperature induced by the 21~cm line {emitted} by a patch
of neutral hydrogen relative to the background radiation and given by
(e.g., \cite{Madau:1996cs,Barkana:2000fd,Furlanetto:2006jb})
\begin{align}
\delta T_b(z) 
&\simeq  9 
(1+\delta_b) 
x_{\mathrm{HI}} 
{(1+z)}^{1/2}
\left[1-\frac{T_\gamma (z)}{T_{\rm S}}\right] 
\left[\frac{H(z)/(1+z)}{dv_\parallel/dr_\parallel}\right]
 \mathrm{mK}\,,
\label{eq:Tb}
\end{align}
where 
$\delta_b$ is the fractional baryon density perturbation,
$x_{\mathrm{HI}}$ is the neutral fraction of hydrogen,
$H(z)$ is the Hubble parameter 
and ${dv_\parallel}/{dr_\parallel}$ is the 
proper velocity gradient along the line of sight.
The spin temperature $T_{\rm S}$ quantifies the relative number 
densities of atoms in the two hyperfine levels of the electronic ground state, and
$T_\gamma$ is the brightness temperature of the background radiation field, 
which is CMB in most cases.
The quantity $\delta_b$ is the tracer of the total matter density, 
which is relevant for 
probing the primordial density 
fluctuations, and quantities $x_{\rm HI}$ and $T_{\rm S}$ are driven by 
the astrophysical processes that model the formation of 
the first luminous sources
and reionization of the universe.

We use \texttt{21cmFAST v2}
\cite{Mesinger:2011,21cmfast}, a 
semi-numerical simulation code,
to simulate the cosmological $21$ cm signal. 
\texttt{21cmFAST} generates realizations of the 
density, ionization,
spin-temperature fields, and 
velocity gradient, which 
can be combined to give  
the 21~cm brightness temperature field using eq.~\eqref{eq:Tb}. 
We work in the framework of
Friedmann-Lema\^{i}tre-Robertson-Walker cosmology
with a flat spatial geometry.
The concordance $\Lambda$CDM model
is parameterized by 
the baryon density 
$\omega_b = \Omega_b h^2$, 
the cold dark matter density 
$\omega_{\text{cdm}} = \Omega_{\text{cdm}} h^2$,
the present Hubble parameter $H_0$,
the optical depth to reionization 
$\tau$, 
the amplitude of the scalar perturbations $A_{\rm s}$ 
at the pivot scale $k_0=0.05({\rm Mpc}^{-1})$
(or $\sigma_8$, the variance of the density fluctuations 
within	a sphere of $8 h^{-1}{\rm Mpc}$ radius) and the spectral index $n_s$.
The brightness temperature of the 21~cm line is sensitive to the 
underlying astrophysical 
{processes,  which in the simulations are modelled as being parametrized by  some variables.} 
Some of the important parameters are: 
\begin{itemize}
\item $M_{\rm min}$ - the halo mass below which the abundance of
active star-forming galaxies is exponentially suppressed 
\item $f_{\rm esc}$ - the normalization of the ionizing UV escape fraction of high-$z$ galaxies 
\item $\alpha_{\rm esc}$ - the power-law scaling of $f_{\rm esc}$ with halo mass 
\item $f_\ast$ - the fraction of galactic gas in stars 
\item $\alpha_\ast$ - the power-law scaling of $f_\ast$ with halo mass 
\item $t_\ast$ - the star formation time-scale taken as a fraction of the Hubble time
\item $E_0$ - the minimum X-ray photon energy capable of escaping
the galaxy and
\item $L_{X<2{\rm keV}}/$SFR - the normalization of the soft-band X-ray
luminosity per unit star formation computed over the band 2 KeV.
\end{itemize}
%
%
%
The astrophysical parameters are 
extremely uncertain at high redshifts. 
Therefore, MCMC sampling of all the uncertain 
parameters is required when the actual data arrive.
However, simultaneous variation of all the parameters becomes
computationally expensive, particularly when the 
initial conditions need to be generated every time to sample
the cosmological parameters.  

To perform the analysis in a reasonable computational time, 
we choose two of the astrophysical parameters
that are important for EoR modelling.
The first is 
$M_{\rm min}$ (in solar mass units $M_\odot$),  
which can be expressed in terms of the virial temperature
$T_{\rm vir}$ (K) as \cite{Barkana:2000fd} 
\begin{equation}
    M_{\rm min}
    =
    10^8 h^{-1}
    \left(\frac{\mu}{0.6}\right)^{-3/2}
    \left(\frac{\Omega_m}{\Omega_m^z} \frac{\Delta_c^z}{18\pi^2}\right)^{-1/2}
    \left(\frac{T_{\rm vir}}{1.98\times 10^4{\rm K}}\right)^{3/2}
    \left(\frac{1+z}{10}\right)^{-3/2}
    { M_\odot}\,,
    \label{eq:Mmin}
\end{equation}
where $\mu$ is the mean molecular weight, 
$\Omega_m^z = \Omega_m (1+z)^3/[\Omega_m(1+z)^3+\Omega_\Lambda]$, and 
$\Delta_c^z=18\pi^2+82d^z-39{d^z}^2$, where $d^z=\Omega_m^z-1$.
The second is the UV ionizing efficiency of high-$z$ galaxies $\zeta$, which is modelled as 
\begin{equation}
\zeta =  30 
\left(\frac{f_{\rm esc}}{0.12}\right)
\left(\frac{f_\ast}{0.05}\right)
\left(\frac{N_\gamma}{4000}\right)
\left(\frac{1.5}{1+n_{\rm rec}}\right)\,,
\label{eq:zeta}
\end{equation}
where 
$N_\gamma$ is the number of ionizing photons produced per baryons in stars
and 
$n_{\rm rec}$ is the typical number of times a hydrogen atom recombines. 
{For high-$z$ galaxies, the parameters
$f_{\rm esc}$ and $f_\ast$ 
are very uncertain.
Keeping in mind a reasonable computational time for the analysis, 
we consider the product of eq~\eqref{eq:zeta} 
as a free parameter rather than varying each of them} 
{individually}. 
Though other parameters such as $L_{X<2{\rm keV}}/{\rm SFR}$ and
$E_0$ contribute to increasing the power spectrum amplitude at 
certain redshifts \cite{Greig:2017jdj}, 
their overall 21~cm power spectrum profiles 
are distinct from those produced by adding bump-like features\footnote{The effects of adding a bump-like feature on
the 21~cm power spectrum are discussed in appendix~\ref{app:degeneracy}.}.
When detailed modelling is avoided, the parameters
$\zeta$ and $T_{\rm vir}$ specify the ionization field 
of a given region \cite{Mesinger:2011}.
Therefore, considering a minimal model, we vary 
$\zeta$ and $T_{\rm vir}$ in our analysis.
%

We fix our {\em fiducial model} as follows: the cosmological parameters are set following the 
best-fit values of \textit{Planck} 2018 results \cite{Akrami:2018odb}\footnote{The best-fit values obtained with the combination of temperature, polarization and lensing data.}:
$\Omega_b h^2 = 0.022$,
$\Omega_{\text{cdm}} h^2 = 0.120$,
$h = 0.6736$
$\tau = 0.054$,
$\sigma_8 = 0.811$ and
$n_s = 0.965$. 
We fix the astrophysical parameters following \cite{Park:2018ljd}:
$M_{\rm min} = 5\times 10^8 M_\odot$ or $\log T_{\rm vir} = 4.69$,
$\zeta = 30$ (i.e., $f_{\rm esc} = 0.1$, $f_\ast = 0.05$ and 
$N_\gamma = 5000$),
$\alpha_\ast = 0.5$,
$\alpha_{\rm esc} = -0.5$,
$t_\ast = 0.5$,
$E_0 = 0.5$ {KeV} and 
$L_{X<2{\rm {KeV}}}/$SFR $ = 10^{40.5}~\text{erg s}^{-1} M_\odot^{-1} {\rm yr}$.

In this work, we  introduce the additional parameters of the bump-like features 
of the primordial power spectrum described in section~\ref{sec:model}. 
In \texttt{21cmFAST}, 
the primordial power spectrum is calculated using the 
default power-law form \cite{Eisenstein:1997ik,Eisenstein:1997jh}, 
{which we modify} according to eq.~\eqref{eq:single_bump}.
{We first study the impact of primordial features on the differential brightness temperature 
while keeping the astrophysical parameters fixed to their fiducial values.
Next, we perform simultaneous sampling of $T_{\rm vir}$ and $\zeta$ to vary 
along with the parameters of primordial features 
to carry out the joint parameter estimation.}
To account for the inaccuracy of semi-numerical simulations, 
we add a modelling uncertainty of $20\%$ in our analysis
(see also section.~\ref{sec:mcmc_analysis}). 
%

\subsubsection{Statistical tools - 21 cm global signal and power spectrum}
The redshifted 21~cm observations provide three-dimensional information
on the neutral hydrogen distribution in the universe. 
{In order to extract physical information from the data,
we can use different statistical tools.}

{The simplest tool is  the  sky-averaged global 21~cm signal, denoted by $\bar{\delta T_b}$.
We will discuss how the global 21~cm signal 
can constrain the 
amplitude of the primordial power spectrum in 
section~\ref{subsec:prior}. }
%
{Since the global 21 cm signal is averaged over the entire sky, we lose the information in the spatial fluctuations of the signal. To extract information from the fluctuations, the most commonly used statistic is  the}  power spectrum {$P_{21}({k})$} given by (see e.g. \cite{Mesinger:2019cosm.book})
\begin{equation}
    \left<\delta_{21}(\vec{k}) \delta_{21}^\ast(\vec{k^{\prime}})\right>
    =
    (2\pi)^3 \delta_D^{(3)}(\vec{k}-\vec{k^\prime})
    P_{21}({k})\,,
    \label{eq:powerspec}
\end{equation}
where $\left<\dots\right>$ denotes 
the ensemble average,
$\delta_{21}(\vec{k})$ is the Fourier transform of the quantity 
$\delta_{21}(\vec{x}) \equiv \left[\delta T_b(\vec{x})-\bar{\delta T_b}\right]/\bar{\delta T_b}$ 
and 
$\delta_D^{(3)}(\Vec{k})$ is the 3D Dirac delta function.
{$P_{21}(k)$ is a function of only the magnitude of the wave-number $k$ due to the assumption of homogeneity and isotropy.} 

The 21~cm power spectrum is sensitive to underlying astrophysical processes at 
different redshifts via eq.~\eqref{eq:Tb}. 
Ideally, the imprints of primordial features are more 
prominent in the 21~cm power spectrum at high redshifts where
fluctuations in $x_{\rm HI}$ and $T_{\rm S}$ are negligible. 
However, observations at such high redshifts become
extremely challenging as the radio sky at 
low frequency is dominated by the foreground emission. 
In our MCMC sampling, we use the simulated 21~cm power spectra 
after the formation of the first stars until the end of reionization 
to probe the imprints of primordial features. 

\subsection{Generating expected 21~cm power spectra}
\label{sec:mock_data}
Observations of 21~cm power spectra will contain
the cosmological 21~cm signal and the noise component  
given by
\begin{equation}
P_{21}^{\rm obs}(k)
=
P_{21}^{\rm signal}(k)
+
P_{21}^{\rm noise}(k)\,,
\label{eq:P21}
\end{equation}
where $P_{21}^{\rm signal}(k)$ is the power spectrum
of cosmological redshifted 21~cm signal {without foreground emission} 
and $P_{21}^{\rm noise}(k)$ is the noise power spectrum expected
from SKA-Low. 
In this section, we describe the generation of both these 
components to use the combined power spectrum as `mock' observations.

\subsubsection{Redshifted 21~cm signal}
The {simulation} parameters relevant 
for generating the redshifted 21~cm signal in \texttt{21cmFAST} are 
the length of the simulation box $L$, 
and the number of cells on each side of the simulation box $N$. 
%
The parameters $L$ and $N$ set the minimum and maximum 
wave-numbers simulated by \texttt{21cmFAST}, respectively. 
Typically, a large simulation box size ($L\ge 250$ Mpc)
is required to simulate cosmic dawn and EoR power 
spectra\footnote{Box length $L<250$ Mpc underestimates 
the large-scale power during the cosmic dawn by 
$7 - 9$\% on average \cite{Kaur:2020}.}.
We choose a box length of $600$ Mpc, starting from a higher resolution box with 
the number of cells $600$, 
sampling down to a lower-resolution box 
with the number of cells $200$. 
Such a simulation gives a resolution of $3$ Mpc. 
We simulate the 21~cm power spectra at various redshifts in the range 
$6 \lesssim z \lesssim 20$\footnote{In principle,
21~cm signals from higher redshifts are 
{more} sensitive to primordial density perturbations (see appendix \ref{app:degeneracy}), 
however, the corresponding lower frequencies are  more strongly  dominated by Galactic foreground (mainly synchrotron) emissions. 
Our choice of redshift range 
gives a compromise between the two.}. 

The input values of the parameters $A_{\rm I }$, 
$k_b$, $\zeta$ and $T_{\rm vir}$,
chosen to create the mock observations, 
are mentioned in section~\ref{sec:results} with their results. 
\subsubsection{Noise power spectra}
\label{sec:noise}
The sensitivity of an interferometer 
to the 21 cm signal depends on 
the thermal noise in the interferometric 
visibilities and the sample variance 
{calculated from the number of independent modes
measured by the interferometer.}
%

{We generate the noise power spectra expected from 
the upcoming telescope SKA-Low \cite{ska}
using the package \texttt{21cmSense}
\cite{Pober2013apJ,Pober:2013jna,21cmsense}.} 
SKA-Low will operate in the frequency range
$50 \lesssim \nu ({\rm MHz}) \lesssim 350$
and is expected to probe the universe in its early stage 
with an order of magnitude better sensitivity than the 
currently operating radio telescopes.
The antenna parameters for SKA-Low are given in table~\ref{tab:SKA1-Low}.
The `core' of SKA-Low consists of short-baseline antennas, 
which are useful for sensitivity calculations \cite{SKA:2018ckk}.
The antenna coordinates in latitude and longitude are available on the SKA 
page \cite{ska-sens}.
\begin{table}[tbp]
    \centering
    \begin{tabular}{|c|c|}
        \hline
        Parameters & Values \\
        \hline
        Number of antennae in the core & 224 \\
        Element size [m] & 38 \\
        Latitude & \ang{26;49;29}S \\
        Longitude & \ang{116;45;52}E \\
        Receiver temperature [K] & 100 \\
        Total observation time [hr] & 1080 \\
        \hline
    \end{tabular}
    \caption{SKA-Low specifications used for producing the noise power spectra.}
    \label{tab:SKA1-Low}
\end{table}

The power spectrum of the thermal noise for a single baseline
is given by~\cite{Parson2012}
\begin{equation}
\Delta_{\rm noise}^2(k) \approx X^2Y \frac{k^3}{2\pi^2} \frac{\Omega}{2t} T_{\rm sys}^2 \,,
\label{eq:thermal_noise}
\end{equation} 
where
$X$ converts angles on the sky to transverse distances, 
$Y$ converts from bandwidth to line-of-sight distance,
$\Omega$(sr) is the solid angle of the 
primary beam of one element,
$T_{\rm sys}$ gives the system temperature, and
the integration time is given by $t$.
The system temperature has two parts: 
the sky and the receiver temperature, 
i.e., $T_{\rm sys} = T_{\rm sky} + T_{\rm rec}$. 
The receiver temperature is set at 100 K, 
and the sky temperature is modelled as \cite{Thomson:2001isra.book}
\begin{equation}
T_{\rm sky} =
60 {\rm K} \left(\frac{300 ~{\rm MHz}}{\nu}\right)^{2.55}\,.
\end{equation} 

The sensitivity estimated in \texttt{21cmSense} depends 
on how the foreground wedge is accounted for
while calculating the noise. 
The `moderate' foreground removal model assumes 
a foreground wedge that extends 0.1~$h$(Mpc$^{-1}$),
 and the `optimistic' foreground
removal model assumes a foreground wedge extending 
to the primary field of view \cite{Pober:2013jna}.
%
\subsection{MCMC sampling}
\label{sec:mcmc_analysis}
We use \texttt{21CMMC}
\cite{Greig:2015qca,Greig:2017jdj,21cmmc},
which efficiently samples  
the parameter space within a Bayesian MCMC framework
based on the \texttt{EMCEE PYTHON} module
\cite{Foreman-mackey:2013}. 
\texttt{21CMMC} uses a modified version of 
\texttt{21cmFAST} to reduce computational time. 
We add the analytical power spectrum template 
given by \eqref{eq:single_bump}
in the initial conditions of \texttt{21CMMC}.
To vary cosmological parameters in \texttt{21CMMC}, 
every time the sampler proposes a new parameter position, 
the initial conditions need to be generated with 
a different random seed which becomes computationally expensive.  
Therefore, we chose the length of the simulation box
to be 250~Mpc and a resolution of 3~Mpc (the same as mock observations)
for the sampled simulations to have a reasonable
computational time and
reasonable accuracy when compared to the mock observations
of higher box lengths. 
In \texttt{21CMMC}, sampling is performed in the region 
$0.1 \lesssim k({\rm Mpc}^{-1}) \lesssim 1$\footnote{The 
lower bound accounts for foreground domination in
the moderate foreground removal model, and 
the upper bound accounts for  shot-noise domination \cite{Greig:2015qca}.}.
In addition to the uncertainties from instrument and 
sample variance, a modelling uncertainty of $20\%$
is included to account for the
inaccuracy of semi-numerical approaches
\cite{Greig:2017jdj,Park:2018ljd}.

\subsubsection{Choice of priors}
\label{subsec:prior}
%
\begin{figure}[!tbp]
	\centering
	\includegraphics[width=0.8\textwidth]{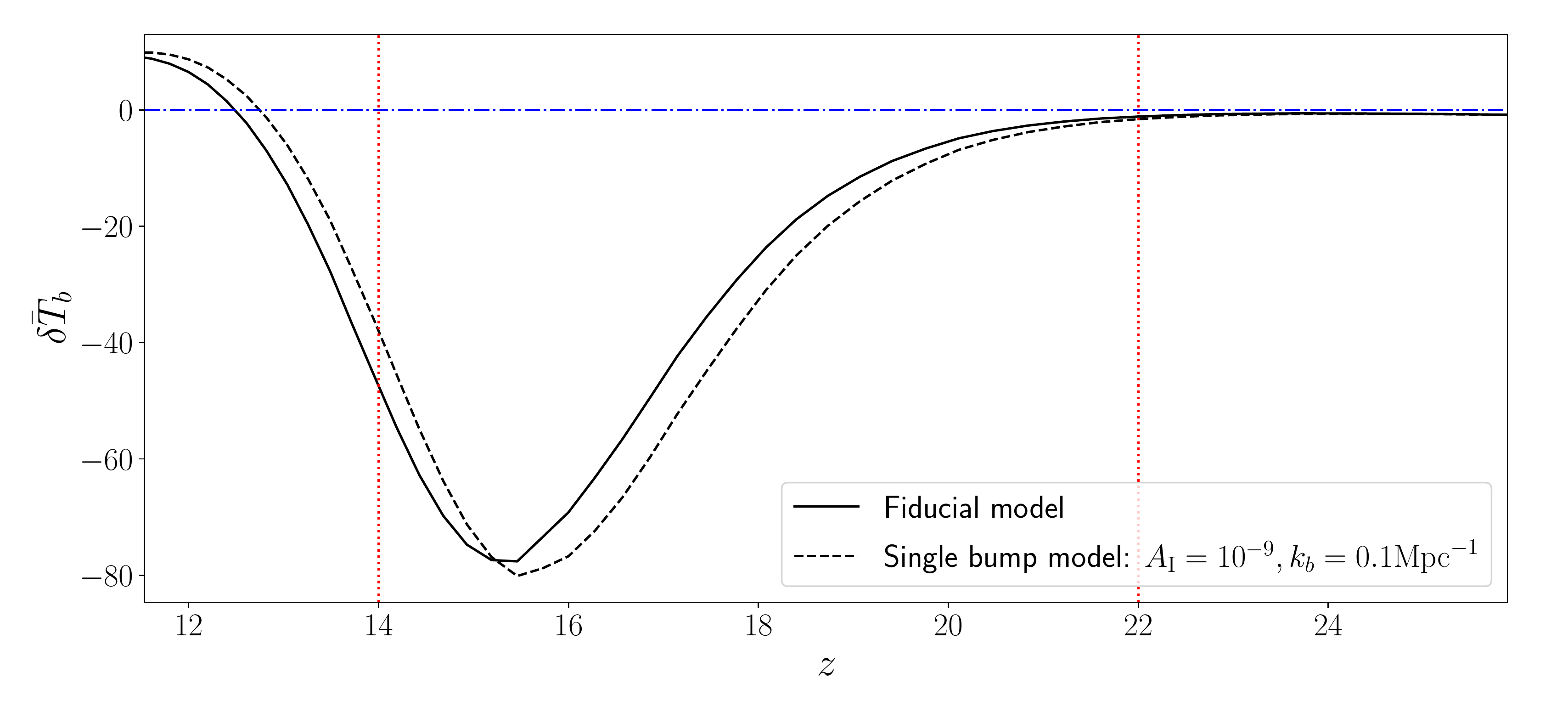}
	\caption{The global 21~cm signals expected for the fiducial and single bump models. 
		The vertical lines indicate the region where the claimed absorption trough of the 
		global 21~cm signal by EDGES \cite{Bowman:2018yin} lies. } 
	\label{fig:global21}
\end{figure}
The single bump model is parameterized by 
the amplitude of the bump $A_{\rm I}$
and the location of the bump $k_b({\rm Mpc}^{-1})$. 
We choose uniform prior probability 
for both parameters
in the Bayesian analysis. 
The choices of upper and lower bounds 
are described in the following.

\paragraph*{Amplitude of the bump:}
The theoretically estimated upper bound on the 
coupling parameter 
is given by {eq.~\eqref{eq:g_upper}}
which puts an upper bound on the 
amplitude via eq.~\eqref{eq:A1} and is given by
$A_{\rm I} \lesssim 10^{-6}$.
However, a bump-like feature located in the 
sampling region  $[0.1, 1]~{\rm Mpc}^{-1}$ and
amplitude $A_{\rm I}>A_s$
increases the power in the lower-$k$ region 
due to the shape of the feature and
affects the parameter $\sigma_8$ or
the variance of density fluctuations. 
We consider an upper bound on $A_{\rm I}$ to be $10^{-9}$
so {that} the value of $\sigma_8$ is consistent with \textit{Planck} 2018 constraints. 
The lower bound on $g^2$ calculated in \cite{Pearce:2017bdc}
is {given by eq.~\eqref{eq:g_lower}},
which gives $A_{\rm I} > 10^{-17}$.

We {first} examine {constraints on $A_{\rm I}$ 
based on already} available 21~cm observations. 
The global 21~cm signal for the fiducial model is plotted in 
figure~\ref{fig:global21}, 
{along with the expected signal for} the single bump model.
The vertical lines indicate the region where the absorption trough 
of the global 21~cm signal {obtained} by the EDGES experiment \cite{Bowman:2018yin} lies.
Various groups have critically examined the EDGES results, 
and one of the recent results from \cite{Singh:2022NatAs}
rejected the best-fit 
profile found by \cite{Bowman:2018yin}
with a $95.3\%$ confidence level.
Figure~\ref{fig:global21} shows that 
when the amplitude of the primordial power spectrum is increased by incorporating a primordial feature, 
the global 21~cm profile shifts towards the higher redshift.
The reason for this shift is that the amplitude of the primordial fluctuations at 
small-scales affects structure formation, 
further affecting the timing when the Lyman-$\alpha$ sources are produced 
and {thereby} changing the redshift evolution of the global 21~cm signal
\cite{Yoshiura:2018zts,Yoshiura:2019zxq}.
The constraints on $A_{\rm I}$ are weaker 
even if we take the EDGES observations into account. 

\begin{figure}[!tbp]
    \centering
    \includegraphics[width=1.05\textwidth]{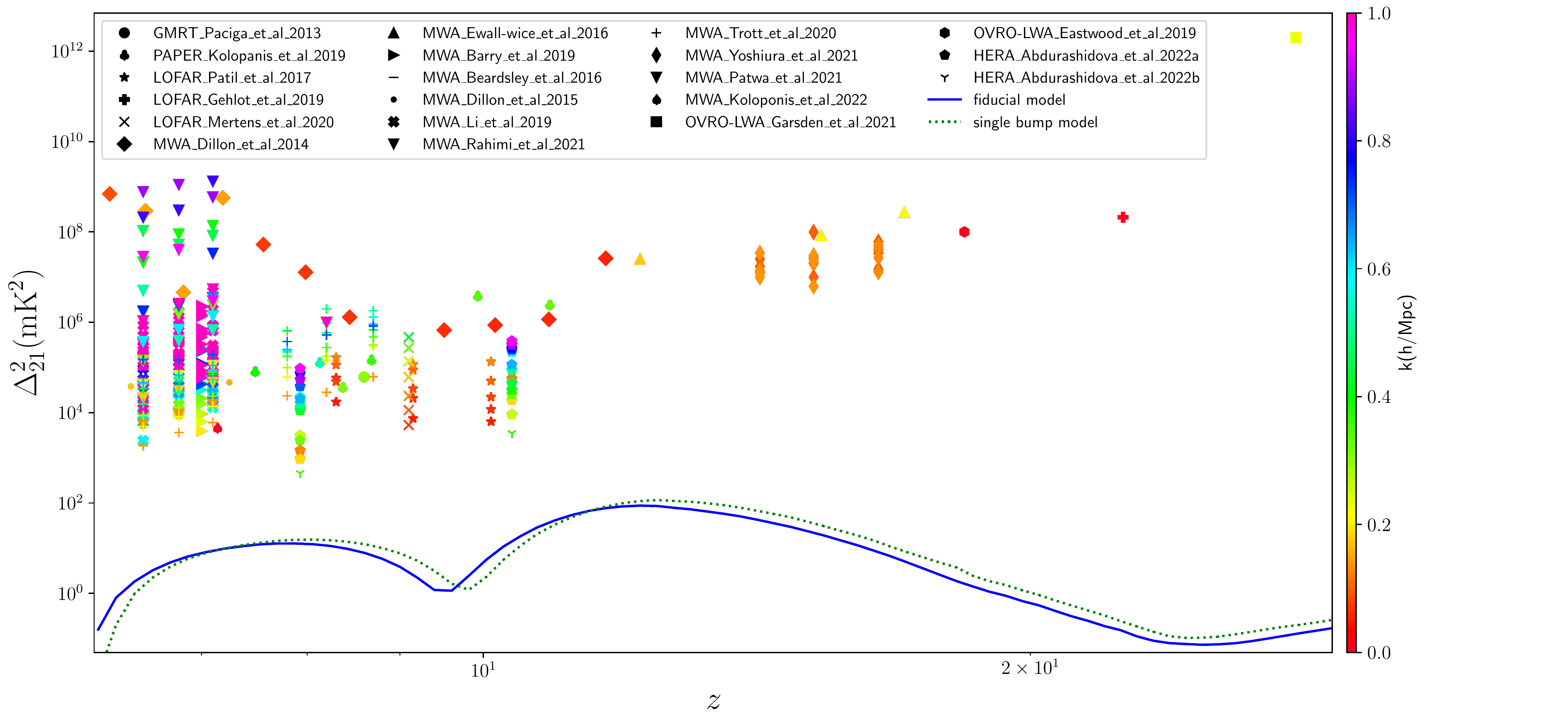}
    \caption{The upper limits on 21~cm power spectra from recent observations 
    	are shown with different markers. 
    	The colour bar gives the co-moving wave-number $k$ in ($h{\rm Mpc}^{-1}$) units.
    	The simulated 21~cm power spectra for the fiducial and single bump model
    	($A_{\rm I}=10^{-9}$, $k_b = 0.1 ({\rm Mpc}^{-1})$ or $k_p \simeq 0.335 ({\rm Mpc}^{-1})$) 
    	are plotted in solid and dotted curves, respectively.
    	The evolution of the simulated power spectra is shown at $k= 0.33 ({\rm Mpc}^{-1})$.
     }
    \label{fig:upper_limit}
\end{figure}

We also investigate the possible constraints from published upper limits
on 21~cm power spectra.  
In figure~\ref{fig:upper_limit}, we plot  
21~cm power spectra in the following form
\begin{equation}
 \Delta_{21}^2({k}) := \left(\frac{k^3}{2\pi^2}\right)P_{21}({k})\,. 
 \label{eq:PS_Delta}
\end{equation}
The observational upper limits plotted are from
facilities such as 
GMRT \cite{Paciga:2013fj}, 
LOFAR \cite{Mertens:2020llj,Patil:2017zqk,LOFAR:Gehlot}, 
MWA \cite{Dillon:2013rfa,Barry:2019qxp,Beardsley:2016,Dillon:2015pfa,Ewall-Wice:2016ysg,Trott:2020szf,Yoshiura:2021yfx,Pober:2019,Patwa:2021dma}, 
PAPER \cite{Kolopanis:2019vbl}, OVRO-LWA \cite{Eastwood:2019rwh,Garsden:2021kdo}, 
AARTFAAC \cite{Gehlot:2020pul} and HERA \cite{HERA:2021bsv,HERA:2022wmy}.
The colour associated with each data point is given by the $k$ value shown in the colour bar.
For comparison with the expected 21~cm power spectra 
from single bump models, 
we also plot the redshift evolution of the corresponding 
21~cm power spectra at $k=k_p=0.33({\rm Mpc}^{-1})$
(the bump was introduced at $k_b = 0.1 ({\rm Mpc}^{-1})$).
Figure~\ref{fig:upper_limit} shows that 
the current upper limits are a few orders of magnitude greater 
than the simulated 21~cm power spectra for the fiducial and single bump models. 
Following the above arguments, 
we {choose} an upper bound on the prior of
$A_{\rm I}$ to be $10^{-9}$. 
The lower bound on $A_{\rm I}$ is set to be $10^{-11}$ to have a distinguishable
effect on the 21~cm power spectra. 
We assume a uniform prior probability distribution on $\log A_{\rm I}$. 

\paragraph*{Location of the bump:}
A bump-like feature at location $k_b$ on
the primordial power spectrum peaks at a higher-$k$ value
given by eq.~\eqref{eq:kpeak}. 
We note that the parameter $k_b$ is not
restricted from the theoretical models. 
In order to make the peak of the bump 
appear in the co-moving wave-number range
$[0.1, 1]~{\rm Mpc}^{-1}$, 
we choose a uniform prior probability for $k_b$
in the limits $[0.0299, 0.299]~{\rm Mpc}^{-1}$.

\section{Results}
\label{sec:results}
This section describes the constraints
on primordial bump-like features 
expected with near-future observations of 21~cm power spectrum 
from SKA-Low. 

We investigate the case of a single burst of particle production
during inflation that predicts a single bump-like feature
on the primordial power spectrum given by eq.~\eqref{eq:single_bump}. 
We first consider a scenario where we fix all the 
astrophysical parameters to their fiducial values
as mentioned in section~\ref{subsec:21cm},
and vary only the single bump model parameters in the
MCMC sampling. 
We then discuss a scenario with the simultaneous variation
of astrophysical and single bump model parameters. 

\subsection{Constraining the single bump model parameters {keeping astrophysical parameters fixed}}
To generate mock 21~cm power spectra, 
we follow the methodology described in section~\ref{sec:mock_data}. 
We choose various input values of the single bump model parameters to 
produce mock power spectra and investigate the ability
of SKA-Low to constrain them. 

\paragraph*{Case I:}
\begin{table}[!tbp]
	\renewcommand{\arraystretch}{1.5}
	\centering
	\resizebox{\textwidth}{!}
	{\begin{tabular}{|c|c|c|c|c|c|c|c|c|c|c|}
	\hline
        \multirow{4}{*}{\bf Parameters} &  \multicolumn{10}{c|}{\bf Single bump models} \\
	\cline{2-11}
	& \multicolumn{5}{c|}{\bf Case I} & \multicolumn{5}{c|}{\bf Case II} \\
	\cline{2-11}
	& \multirow{2}{*}{\bf Input} &  \multicolumn{2}{c|}{\bf Optimistic} & \multicolumn{2}{c|}{\bf Moderate} 
	& \multirow{2}{*}{\bf Input} &  \multicolumn{2}{c|}{\bf Optimistic} & \multicolumn{2}{c|}{\bf Moderate}\\
	\cline{3-6}
	\cline{8-11}
	& & Median & $\delta_\theta$(\%) & Median & $\delta_\theta$(\%)& & Median & $\delta_\theta$(\%)& Median &$\delta_\theta$(\%)\\
	\hline
        $\log A_{\rm I}$ &  $-9.0$ & $-9.0619^{+0.1413}_{-0.1645}$ &  {1.68} 
        & $-9.0620^{+0.1414}_{-0.1646} $ & {1.68}
        & $-9.3010$ & $-9.3110^{+0.1165}_{-0.1134} $  & {1.23}
	& $-9.3077^{+0.1248}_{-0.1185} $ & {1.31}\\ 
        $k_b({\rm Mpc}^{-1})$ &  $0.1$ & $0.0978^{+0.0125}_{-0.0139} $ & {13.49} 
        & $0.0978^{+0.0126}_{-0.0140}$ & {13.59} 
	& $0.07$ & $0.0688^{+0.0098}_{-0.0115} $ & {15.48} 
	& $0.0692^{+0.0099}_{-0.0121} $ & {15.89} \\
        \hline  \hline
        \multirow{3}{*}{\bf Parameters} & \multicolumn{5}{c|}{\bf Case III} & \multicolumn{5}{c|}{\bf Case IV} \\
	\cline{2-11}
	& \multirow{2}{*}{\bf Input} & \multicolumn{2}{c|}{\bf Optimistic} & \multicolumn{2}{c|}{\bf Moderate}
	& \multirow{2}{*}{\bf Input} & \multicolumn{2}{c|}{\bf Optimistic} & \multicolumn{2}{c|}{\bf Moderate}\\
	\cline{3-6}
	\cline{8-11}
	& & Median & $\delta_\theta$(\%) & Median & $\delta_\theta$(\%)& & Median & $\delta_\theta$(\%)& Median &$\delta_\theta$(\%)\\
	\hline
	$\log A_{\rm I}$ & $-10.3010$ &  $-10.5259^{+0.5220}_{-0.5260} $ & {4.97}
	& $-10.6314^{+0.5418}_{-0.5301} $ & {5.04} 
	& $-10.5228$ & $-10.5650^{+0.5420}_{-0.5460} $ & {5.14} 
        & $-10.6240^{+0.5232}_{-0.5746} $ & {5.16}\\ 
	$k_b({\rm Mpc}^{-1})$ & $0.07$ & $0.1088^{+0.0873}_{-0.0579} $ & {66.72}
	& $0.1246^{+0.0963}_{-0.0692} $ & {66.41}
	& $0.04$ & $0.1262^{+0.0922}_{-0.0711} $ & {64.69}
        & $0.1262^{+0.0963}_{-0.0741} $ & {67.51}\\ 
        \hline
        \end{tabular}}
        \caption{The results of single bump models investigated. 
            For each case, we quote the input values used 
		to produce the noise power spectra,  
		the recovered median values from one-dimensional 
		marginalized PDFs (with $68\%$ CL) from SKA-Low 
            and {fractional uncertainties as given in eq.~\eqref{eq:fract}}.}
    \label{tab:singlebump_results}
\end{table}

We first consider a single bump model with $A_{\rm I} = 10^{-9}$, 
or about $50\%$ of $A_s$
and $k_b = 0.1 ({\rm Mpc}^{-1})$, 
which peaks at ${k_p}\sim 0.33 ({\rm Mpc}^{-1})$.
%
We use both optimistic and moderate foreground removal models
to generate the sensitivity expected from SKA-Low
and consider a modelling uncertainty of 20\% to create the 
noise power spectra. 
Table~\ref{tab:singlebump_results} 
provides 
the input values used to produce the noise power spectra and 
the recovered median values of the parameters 
obtained from the one-dimensional marginalized 
Probability Distribution Functions (PDF)
with 
a $68\%$ Confidence Level (CL).
{We also quote the fractional uncertainty on each recovered parameter $\theta$
having uncertainty $\sigma_\theta$
($68\%$ CL) as}
\begin{equation}
    \delta_\theta := \frac{\sigma_\theta}{\theta}\,.
    \label{eq:fract}
\end{equation}
The two-dimensional posterior PDFs 
with the marginalized one-dimensional PDFs
of single bump model parameters are plotted in 
{the top row of} figure~\ref{fig:SB_results}\footnote{{The
triangle plots of posterior PDFs are generated with the package
\texttt{GetDist} \cite{Lewis:2019xzd}.}}.
{Our results indicate that a primordial bump-like feature
with parameters $A_{\rm I} \sim 0.5 A_s$
and $k_b = 0.1 ({\rm Mpc}^{-1})$
are recoverable} within
$68\%$ CL using the SKA-Low configurations
with an optimistic foreground removal model. 
{We also carry out this analysis with a moderate
foreground removal model, and the changes in the 
recovered values with their uncertainties are negligible,
as reported in table~\ref{tab:singlebump_results}}. 

\paragraph*{Case II:}
Next, we test the detectability of a primordial bump-like
feature with a smaller
amplitude than 
the one mentioned above.
Since the thermal uncertainties estimated from SKA-Low are less
at lower-$k$ values ($k < 0.3 ({\rm Mpc}^{-1})$)
as in eq.~\eqref{eq:thermal_noise},
we create mock power spectra for a single bump model
with the following parameter values: 
$A_{\rm I} = 5\times 10^{-10}$, or about $25\%$ of $A_s$ and
$k_b = 0.07 ({\rm Mpc}^{-1})$, or $k_p \simeq 0.23 ({\rm Mpc}^{-1})$. 
We carry out MCMC sampling for the single bump case
using the noise power spectra
produced with {both} the moderate and optimistic foreground 
removal models. 
{In both cases, $20\%$ modelling uncertainty is added to the noise power, like in the previous case.}
Our results are given in table~\ref{tab:singlebump_results}. 
The one- and two-dimensional PDFs of the parameters 
are shown in the second row of figure~\ref{fig:SB_results},
which clearly indicates recovery of both parameters 
within $68\%$ CL. 
%
Compared with the optimistic case, 
the moderate foreground removal model
increases the fractional errors $\delta_\theta$
(eq.~\eqref{eq:fract}) on 
$\log A_{\rm I}$ and $k_b$ by
$0.08\%$ and $0.4\%$, respectively. 
We emphasize that 
even for the moderate foreground removal scenario,
the parameters of the single bump model
are recovered within $68\%$ CL. 

\paragraph*{Case III:} To test the recoverability of the lowest amplitude of 
the bump-like feature, we perform another MCMC sampling 
with an order of magnitude smaller amplitude than the 
previous case and $k_b$ being the same as the previous case, 
i.e., $A_{\rm I} = 5 \times 10^{-11}$ (or about $2.5\%$ of $A_s$)
and $k_b = 0.07 ({\rm Mpc}^{-1})$.
The input values of the parameters with their recovered values
are given in table~\ref{tab:singlebump_results}.
The corresponding one- and two-dimensional marginalized 
posterior distributions are shown in the third row of figure~\ref{fig:SB_results}. 
Compared to the previous case in which bump-like features at the same $k_b$
with an order of magnitude larger $A_{\rm I}$, 
we found that the fractional errors on $\log A_{\rm I}$
and $k_b$ have increased by about
$4\%$ and $51\%$, respectively. 
Though the result indicates recovery of parameters within $68\%$ CL,
the posterior PDFs show that
the primordial bump-like features of smaller amplitudes
have wider PDFs with large uncertainties. 
The median values from the one-dimensional marginalized distributions 
are away from the input values due to their asymmetric PDFs. 
{However, note that the values 
{corresponding to the maxima of 2D PDFs}
($(\log A_{\rm I}, k_b)_{\rm 2D}= (-10.44, 0.076)$ for the optimistic case
and $(\log A_{\rm I}, k_b)_{\rm 2D}= (-10.28, 0.09)$ for the moderate case, 
{marked with `$\times$' in the figure})
are close to the input values, as seen in the two-dimensional PDFs. }

\paragraph*{Case IV:}
In 
{the current work, MCMC sampling is  being performed 
in the wave-number range $0.1 \lesssim k({\rm Mpc}^{-1}) \lesssim 1$. In \cite{Naik:2022mxn}, 
the wave-number range $0.0002 \lesssim k({\rm Mpc}^{-1}) \lesssim 0.15$ was used 
to obtain upper bounds on $A_{\rm I}$ ($95\%$ CL)
using the \textit{Planck} 2018 data. 
The two ranges have a small overlap region, and the question arises whether SKA-Low
can recover the bump-like feature that was allowed by CMB data.}  
With this motivation, we choose $A_{\rm I}= 3\times 10^{-11}$
and $k_b=0.04({\rm Mpc}^{-1})$ so that the peak of the bump
appears at $k_p\simeq 0.13({\rm Mpc}^{-1})$.
The results are provided in table~\ref{tab:singlebump_results}.
{The one- and two-dimensional PDFs 
are shown in the bottom row of figure~\ref{fig:SB_results}.}
The PDFs indicate that the median of $A_{\rm I}$
is close to the mock value chosen and recovered within 
$68\%$ CL. However, $k_b$ is recovered, 
not within 68\%, but within $95\%$ CL.  
We conclude that as the amplitude of the feature 
decreases, constraining $k_b$  
becomes more difficult with only a two-point correlation function. 

{We summarize the results of all the above cases in figure~\ref{fig:all_results}, 
which capture the trend of the parameter estimation, 
and the variation of the uncertainties as the true value of $A_{\rm I}$ is made smaller. 
We reiterate the important points that can also be discerned visually from this figure. 
First, the errors increase mildly from the optimistic to the moderate foreground removal scenario. 
{Secondly, as the errors on 21~cm power spectra are scale-dependent, 
Case II, which has input values of both parameters smaller than Case I, 
has marginally smaller fractional uncertainty for $A_{\rm I}$,} and the bias of the recovered versus input parameters values is also smaller. Thirdly, as we decrease $A_{\rm I}$ further, the errors increase, and the recovered median  values of $A_{\rm I}$ show mild bias (within 68\% CL) towards the lower side, while the recovered median $k_b$ values show relatively stronger bias (beyond 68\% CL) towards the higher side. \ 
{The bias becomes stronger as the input value of $A_{\rm I}$ is decreased. This trend implies that $k_b$ becomes harder to constrain.} 

\begin{figure}[!tbp]
	\centering
	\fbox{\bf Case I}\\
	\includegraphics[height=0.205\textheight,width=0.41\textwidth]{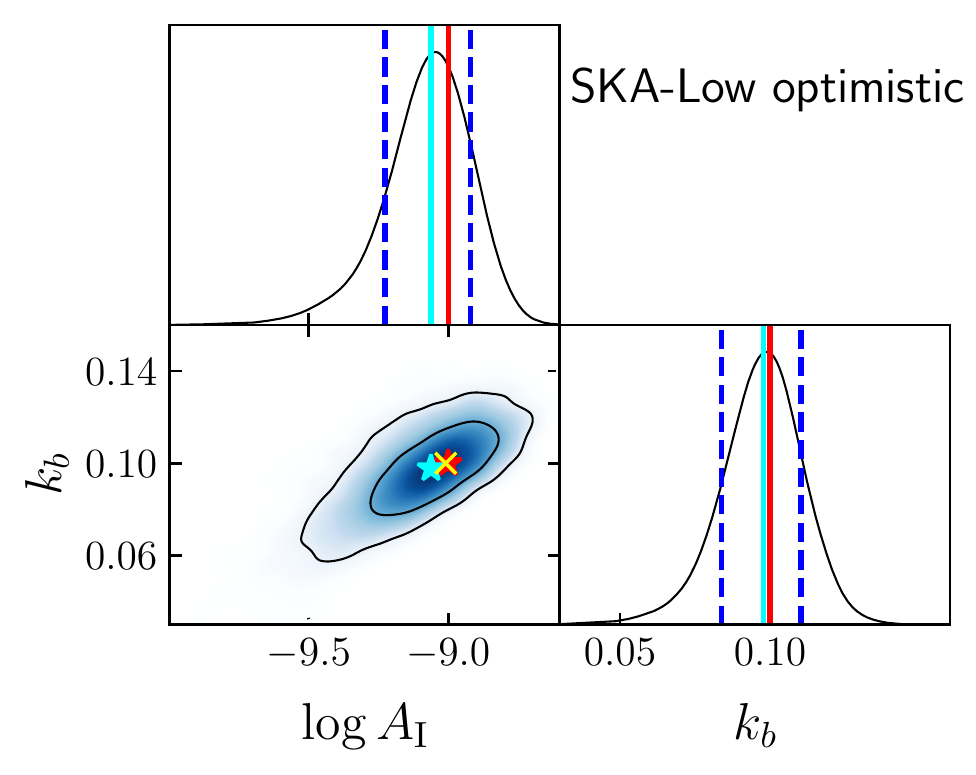}
	\includegraphics[height=0.205\textheight,width=0.41\textwidth]{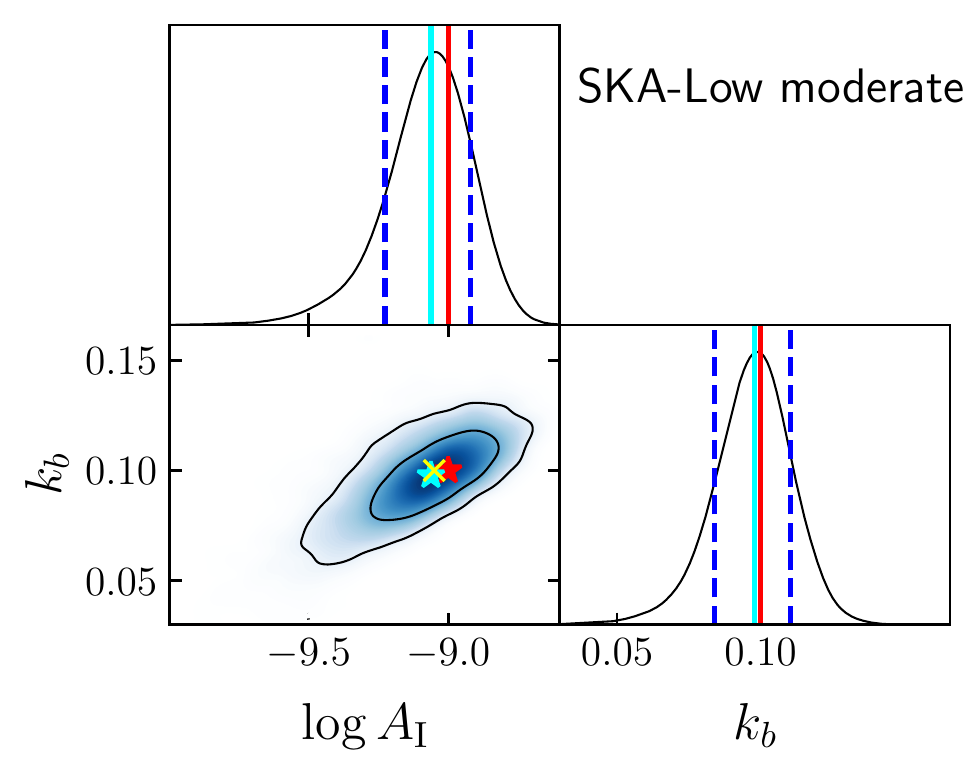}\\
	\fbox{\bf Case II}\\
	\includegraphics[height=0.205\textheight,width=0.41\textwidth]{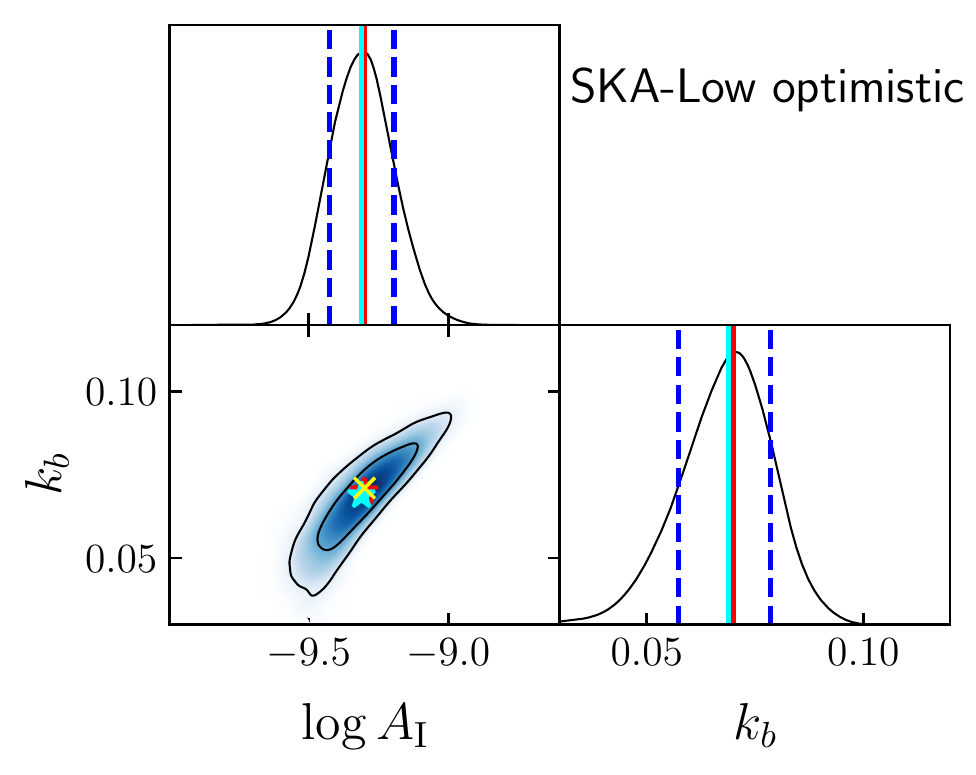}
	\includegraphics[height=0.205\textheight,width=0.41\textwidth]{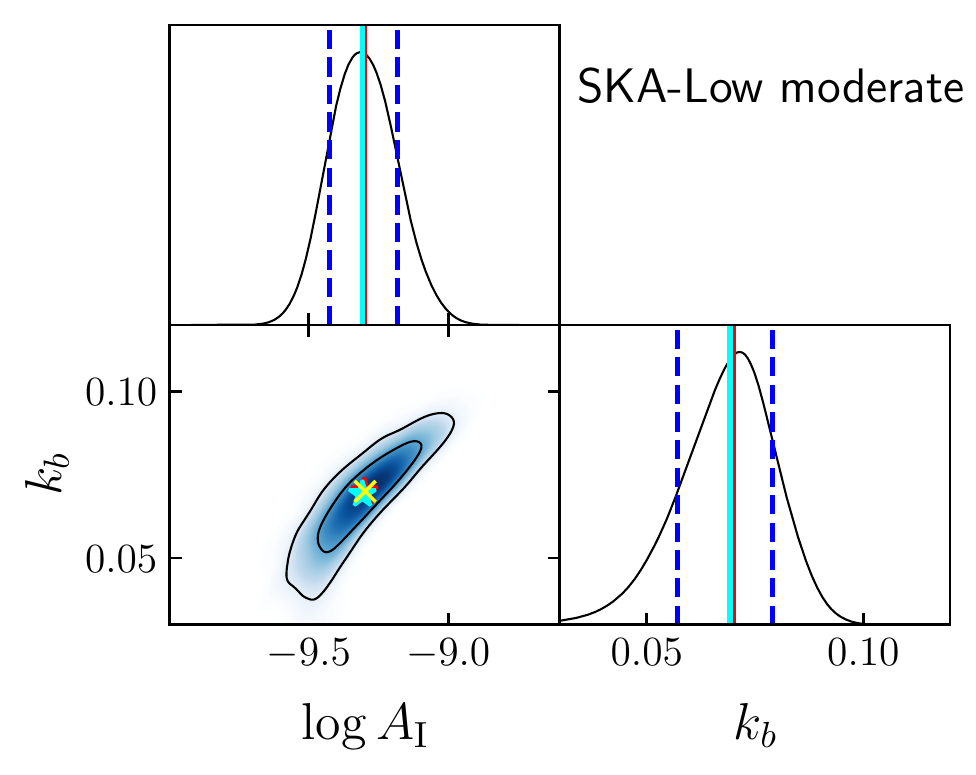}\\
	\fbox{\bf Case III}\\
	\includegraphics[height=0.205\textheight,width=0.41\textwidth]{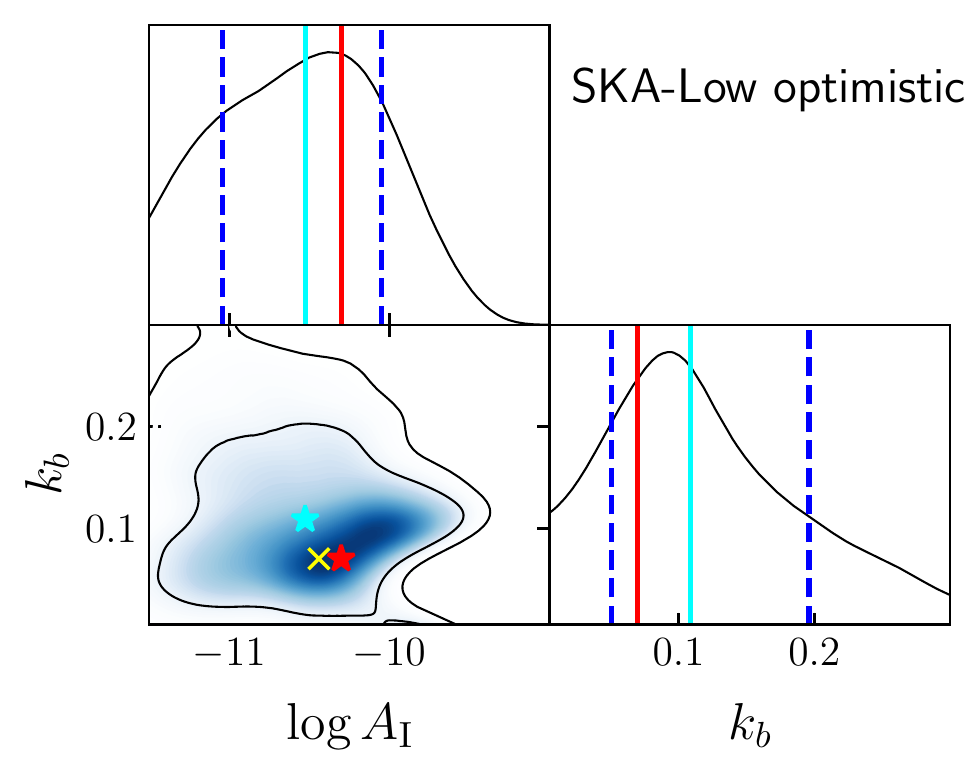}
	\includegraphics[height=0.205\textheight,width=0.41\textwidth]{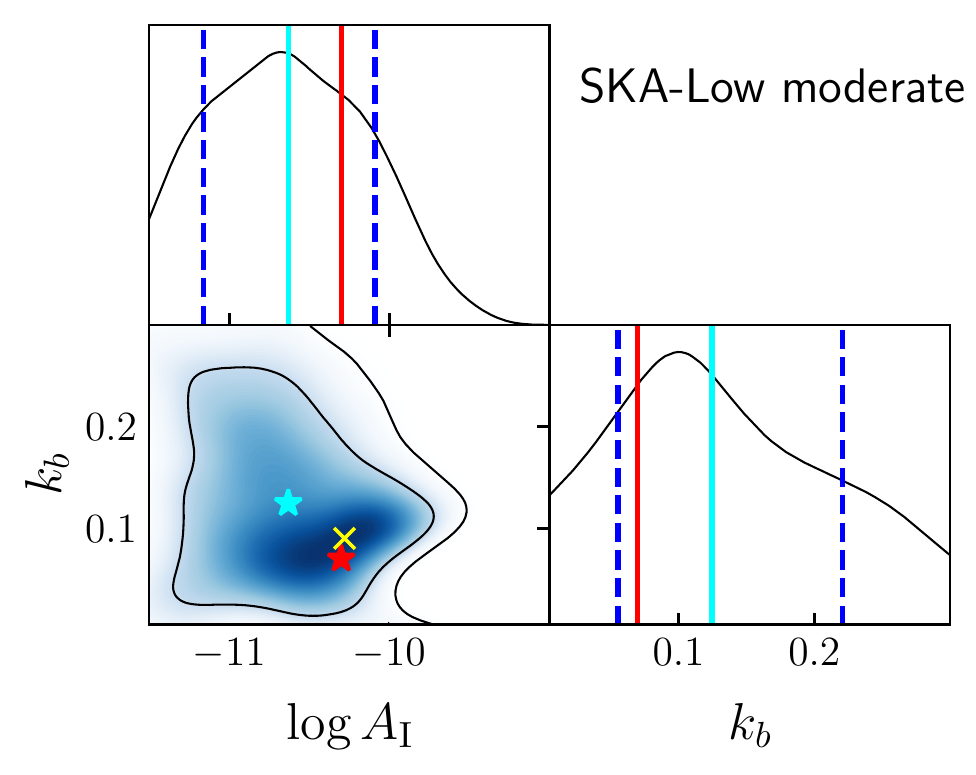}\\
	\fbox{\bf Case IV}\\
	\includegraphics[height=0.205\textheight,width=0.41\textwidth]{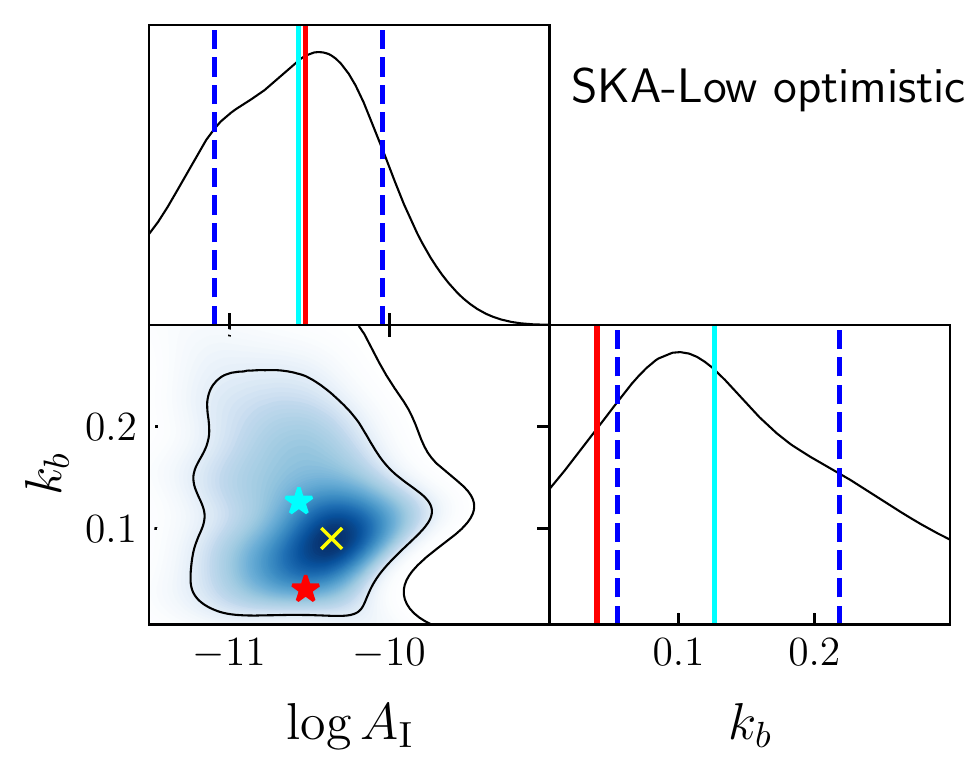}
	\includegraphics[height=0.205\textheight,width=0.41\textwidth]{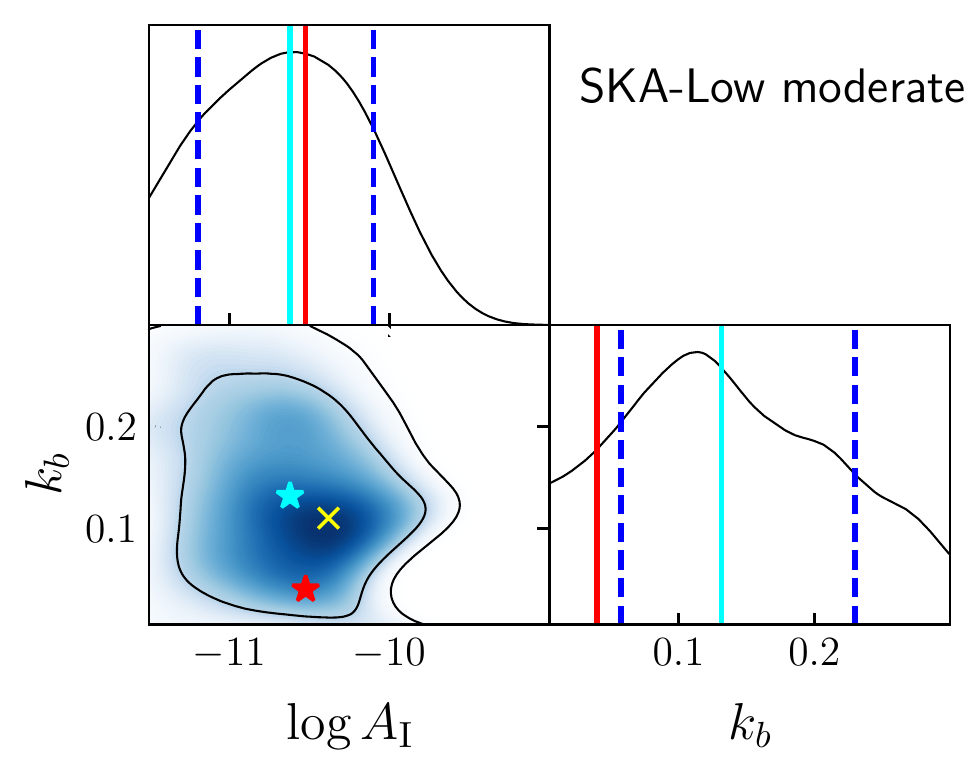}
	\caption{\label{fig:SB_results}\small{Posterior PDFs for the single bump model parameters $\log A_{\rm I}$ and $k_b$
			using the sensitivity from SKA-Low with
			optimistic ({\em left}) and moderate ({\em right})
			foreground removal models. 
			The two contours indicate the
			parameter space corresponding to $68\%$ and
			$95\%$ CL, respectively.
			The red (cyan) 
			{star and lines mark the} input parameter values (recovered median values from one-dimensional PDFs)
			(table~\ref{tab:singlebump_results}).
			The vertical dashed lines indicate $68\%$ 
			parameter space enclosed in one-dimensional PDFs.
			{The markers `$\times$' denote the maxima in each of the two-dimensional PDFs.}}}
\end{figure}
\begin{figure}[tbp]
	\centering
\includegraphics[width=0.49\textwidth]{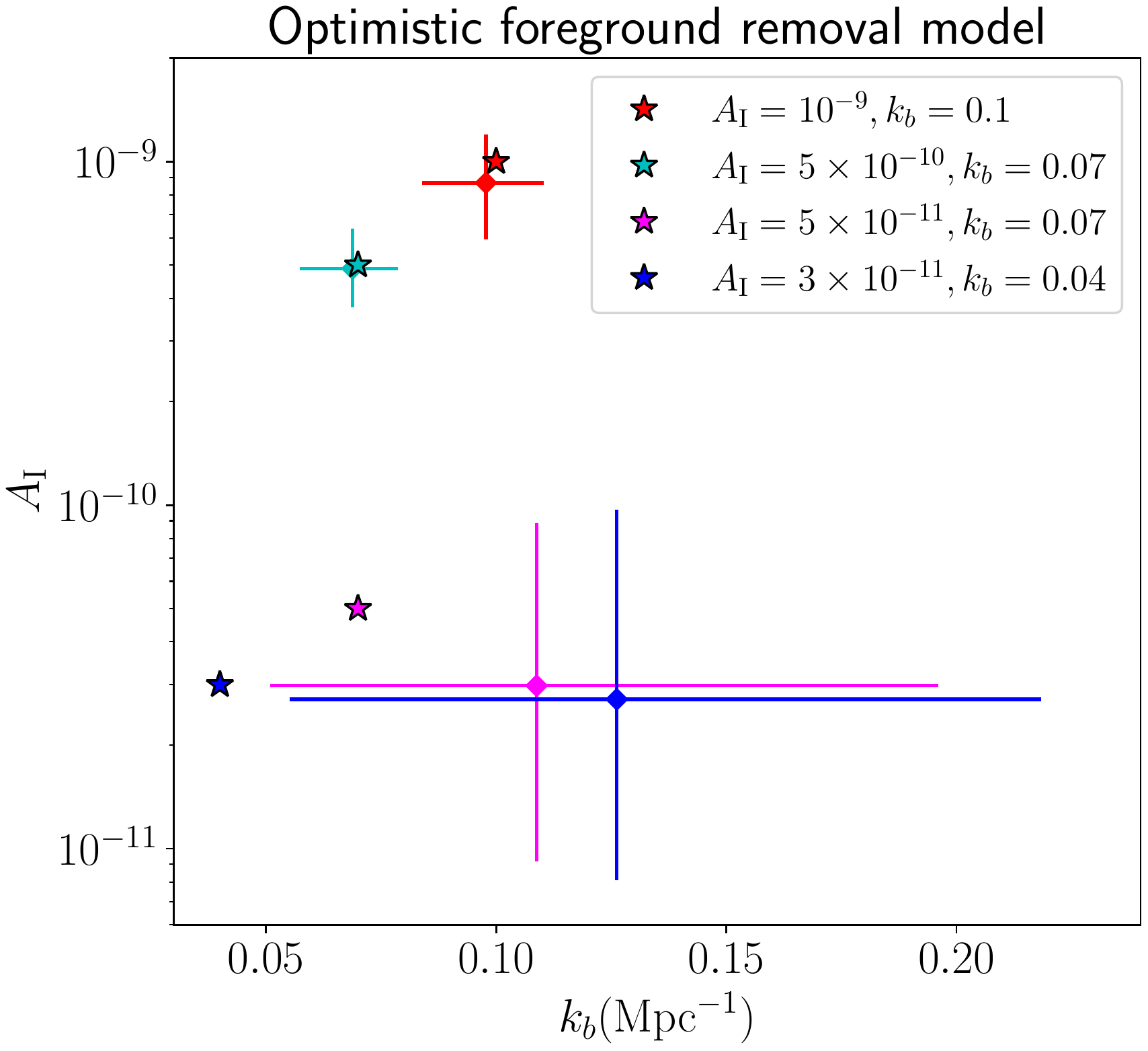}         
\includegraphics[width=0.49\textwidth]{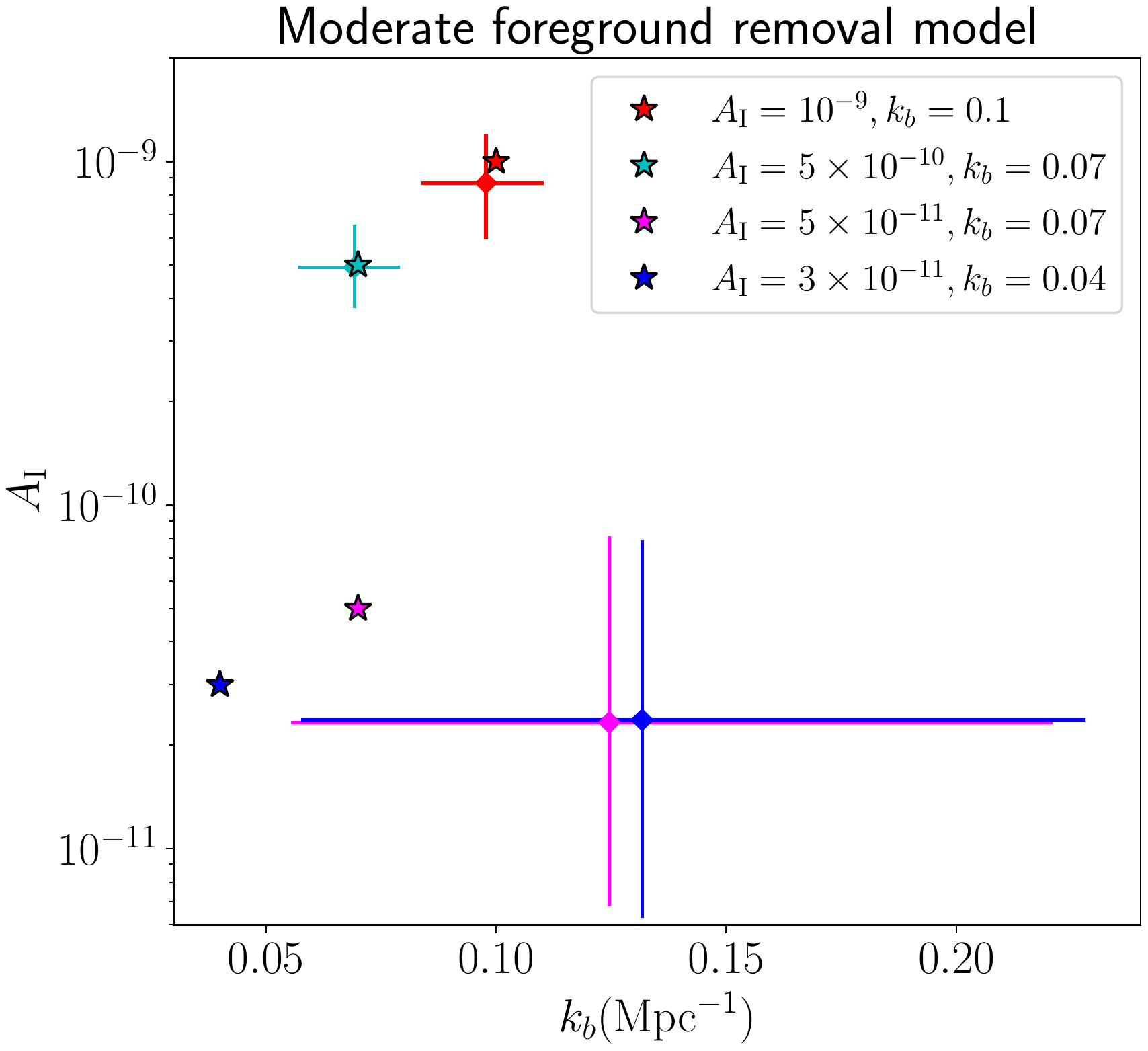}
	\caption{The median values of $A_{\rm I}$ and $k_b$ with 68\% CL uncertainties  obtained from the analysis of single bump models with optimistic ({\em left}) and moderate ({\em right}) foreground removal models. The input values are represented with stars. Colour represents the model investigated. }
\label{fig:all_results}
\end{figure}

{In the above analyses, a modelling uncertainty of 20\% was included. 
Thus, the total error on the 21~cm power spectrum to create the mock observations
is the combination of the sample variance and 
the thermal noise generated from SKA-Low and 
the modelling uncertainty. 
Since the generated sample and thermal noise depend on 
the foreground removal model used, 
one can expect a considerable improvement in the fractional uncertainty 
of the recovered parameters when optimistic foreground removal model is used. 
In figure~\ref{fig:errors}, we show the contribution of these errors
as a function of redshift. 
The errors were generated for the single bump model 
($A_{\rm I}=5\times 10^{-10}$ and $k_b = 0.07\,({\rm Mpc}^{-1})$
or $k_p \simeq 0.235\,({\rm Mpc}^{-1})$) and plotted at $k=k_p$. 
The figure shows that the modelling uncertainty dominates 
most of the redshifts over the other errors generated 
from the optimistic and moderate foreground removal models. 
Hence, in our results, the fractional uncertainties on the recovered 
parameters from the optimistic foreground removal model do not differ
significantly from the moderate case. }
        \begin{figure}[!tbp]
            \centering
            \includegraphics[width=\textwidth]{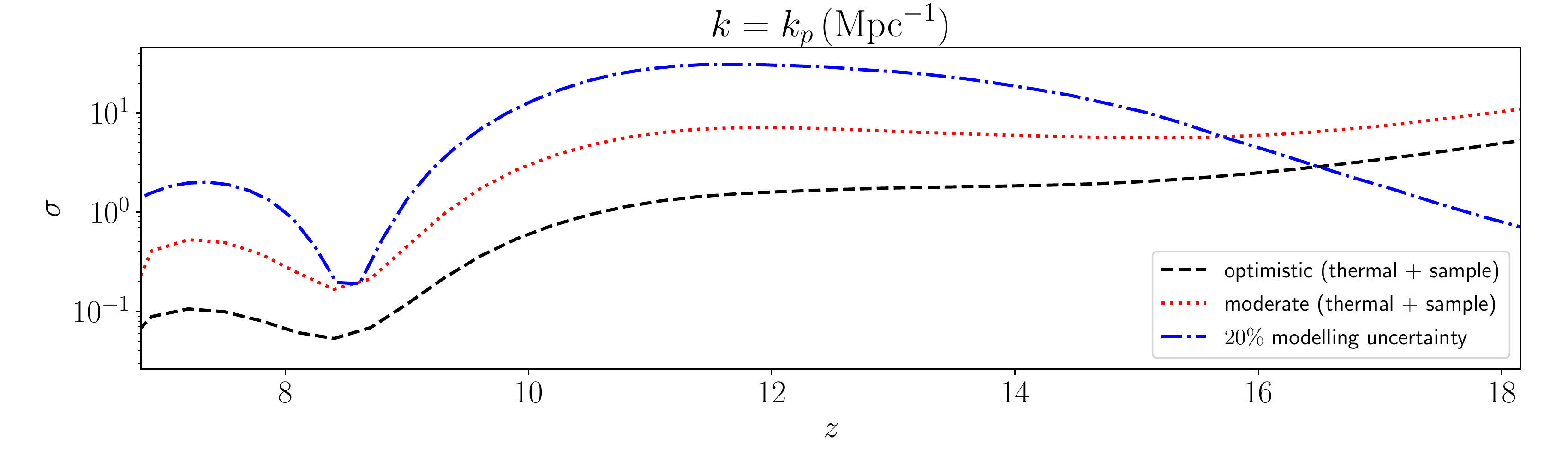}
            \caption{\label{fig:errors} The errors on 21~cm power spectra due to thermal uncertainty
            and sample variance with an optimistic foreground removal model (dashed), 
            moderate foreground removal model (dotted), and
            the modelling uncertainty (dashdotted) plotted at $k=k_p\,({\rm Mpc}^{-1})$.} 
        \end{figure}
In the subsequent section, we also vary two of the astrophysical parameters. 

\subsection{Constraining the parameters of single bump and EoR models simultaneously}
\label{subsec:result_EoR}
%
\begin{table}[tbp]
	\renewcommand{\arraystretch}{1.5}
	\centering
	\begin{tabular}{|c|c|c|c|}
		\hline
		\multirow{2}{*}{\bf Parameters} 
		& \multirow{2}{*}{\bf Input} & \multicolumn{2}{c|}{\bf Recovered values from SKA-Low} \\
            \cline{3-4}
		& & {Median} & {$\delta_\theta$ (\%)}\\
		\hline
		$\zeta$ 
		& $30$  & $30.3317^{+1.4173}_{-1.4583}$ & 4.74\\
		$\log T_{\rm vir}$ 
		& $4.69897$ & $4.8319^{+0.0278}_{-0.0665}$ & 0.97 \\
		$\log A_{\rm I}$ 
		& $-9.3010$  & $-10.1103^{+0.6024}_{-0.5262}$ & 5.58 \\
		$k_b({\rm Mpc}^{-1})$ 
		& $0.07$  & $0.0972^{+0.1027}_{-0.0370}$ & 71.86\\
		\hline
	\end{tabular}
	\caption{The results for the simultaneous constraining of 
            the EoR and single bump model parameters. 
		We quote the input values used 
		to produce the noise power spectra,
		the recovered median values (with $68\%$ CL) 
            from one-dimensional marginalized distributions with SKA-Low sensitivity 
            and the fractional uncertainties as given in eq.~\eqref{eq:fract}.}
	\label{tab:singlebump_EoR_ska}
\end{table}
As mentioned in section~\ref{subsec:21cm}, 
the astrophysical parameters play a significant role 
in modelling the timing of 
the formation of the first luminous sources, 
heating of the IGM, and reionization of the universe.
We consider a minimal model for reionization parameterized by
the 
{log of the} virial temperature $T_{\rm vir}$ (K) 
i.e., $\log T_{\rm vir}$ and the ionizing efficiency $\zeta$.
A uniform prior range for $\log T_{\rm vir}$ 
was assumed in \texttt{21CMMC} is [$4,~6$]. 
The lower limit is motivated by the minimum temperature for 
efficient atomic line cooling, and the  
upper limit is consistent with the host halo masses of 
observed galaxies at $z\sim 6-8$ \cite{Greig:2017jdj}.
A uniform prior range of [10, 250] was chosen for $\zeta$
following \cite{Greig:2017jdj} to explore the models 
where bright, rare galaxies drive reionization.
{For generating the mock data, we choose the values} 
 $\log T_{\rm vir} = 4.69$, consistent with $M_{\rm min} =  5 \times 10^8 ~{M}_\odot$
(following eq.~\eqref{eq:Mmin}), 
and $\zeta = 30$.
\begin{figure}[tbp]
	\centering
	\includegraphics[height=0.6\textheight,width=0.9\textwidth]{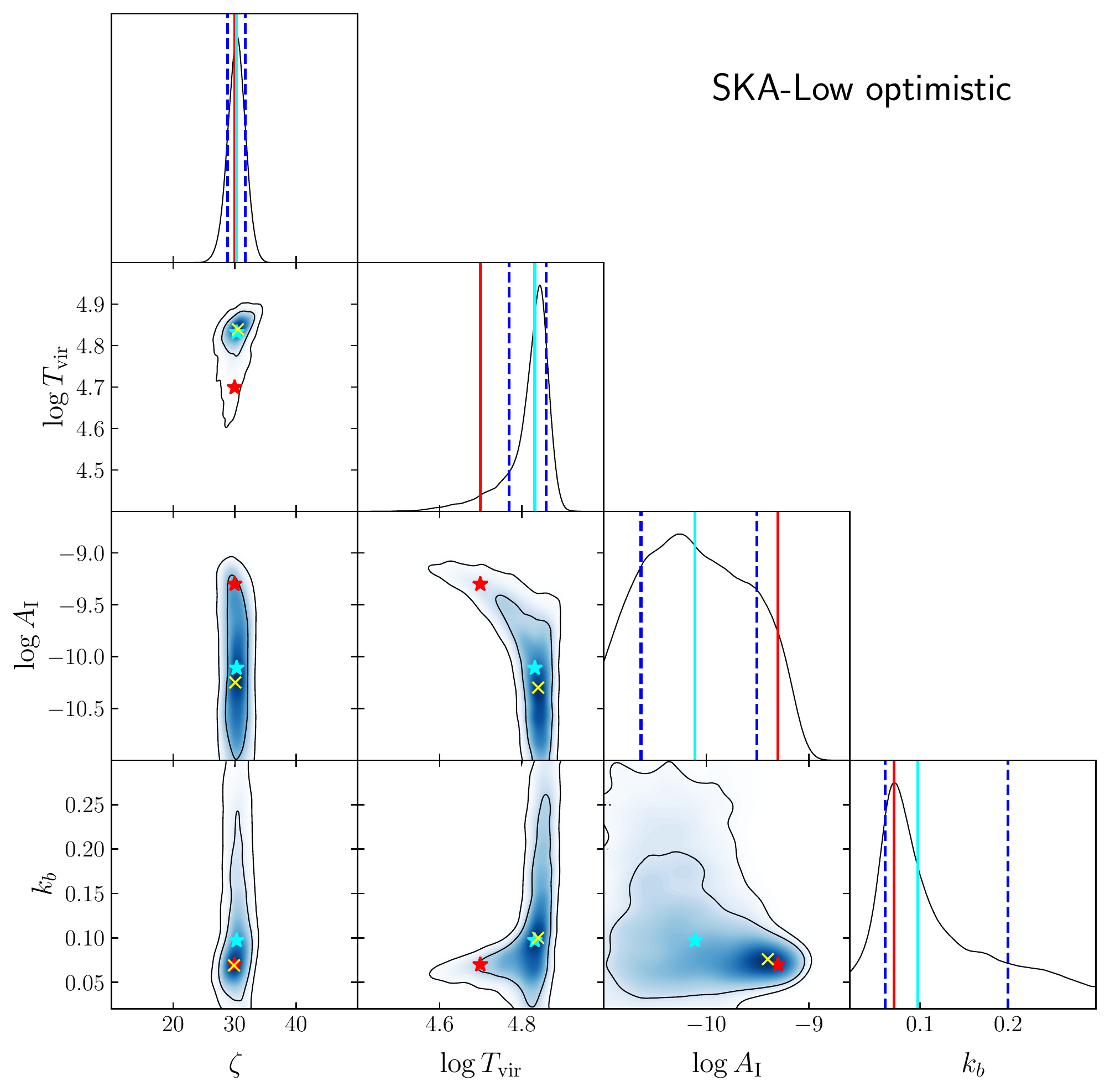}
	\caption{The posterior PDFs for the EoR and single bump model parameters: $\zeta$, $\log T_{\rm vir}$,
	    $\log A_{\rm I}$ and $k_b$. 
		The two contours indicate the
		parameter space corresponding to $68\%$ and
		$95\%$ confidence levels.
		The red (cyan) 
		{star and lines mark the} input parameter values (recovered median values from the SKA-Low)
		(table~\ref{tab:singlebump_EoR_ska}).
		The vertical dashed lines indicate $68\%$ 
	    parameter space enclosed in the one-dimensional PDFs.
        {The markers `$\times$' denote maxima in each two-dimensional PDFs.}
        }
	\label{fig:posterior_singlebump_EoR_SKA}
\end{figure}
\paragraph*{Note on degeneracy:}
A higher value of the parameter $\zeta$
corresponds to more efficient production of ionizing photons
and results in earlier reionization. 
Similarly, a lower value of $T_{\rm vir}$ results in 
earlier reionization as more galaxies can form stars
\cite{Furlanetto:2006jb}. 
As discussed in section~\ref{subsec:prior}, 
the presence of a bump in the primordial power spectrum 
may affect structure formation and consequently change the timings of the epoch of heating and reionization. 
Therefore,  degeneracy between these three parameters - {$T_{\rm vir},\,\zeta$ and $A_{\rm I}$},
is expected. 
{In appendix~\ref{app:degeneracy}, we include a discussion on the effect of the bump-like feature 
on the shape and amplitude of the 21 cm power spectrum
and a comparison with the effect of varying the astrophysical parameters. 
We show that the effects are  distinct from each other.} 

We now simultaneously constrain the EoR parameters 
$\zeta$ and $T_{\rm vir}$ 
and single bump model parameters $A_{\rm I}$ and $k_b$. 
The uncertainties we present are from the
SKA-Low (optimistic foreground removal model) combined with $20\%$
modelling uncertainty. 
Table~\ref{tab:singlebump_EoR_ska} shows the 
parameters,  
their input values {used in} creating mock data, 
and the {recovered median values 
from one-dimensional marginalized posterior PDF.}
In figure~\ref{fig:posterior_singlebump_EoR_SKA},
we show the one- and two-dimensional posterior PDFs of all the 
parameters.
{We can observe that the one-dimensional PDFs of all the parameters, 
	except for $\zeta$, 
	deviate from the Gaussian shape. 
	This reflects the degeneracy among the parameters, which is 
	quite evident from the two-dimensional PDFs.} 
Despite this, our results indicate that
the input parameters are recovered in the following way:
$\zeta$ and $k_b$ within $68\%$ CL, and 
$\log T_{\rm vir}$
and $\log A_{\rm I}$ within $95\%$ CL. 
{It is worth mentioning that  the 
{parameter values corresponding to 
a maximum in the two-dimensional PDF for $\log A_{\rm I}$ - $k_b$
 (i.e., $(\log A_{\rm I}, k_b)_{\rm 2D} = (-9.41, 0.076)$)}  
are closer to the input values compared to the recovered median values.}

When we compare the results obtained here with the results for 
fixed $\zeta$ and $T_{\rm vir}$ (Case~II, optimistic),
we find that keeping $\zeta$  and $T_{\rm vir}$ as free parameters 
increases the fractional errors $\delta_\theta$ (eq.~\eqref{eq:fract}) 
on $\log A_{\rm I}$ by $4\%$ and $k_b$ by $56\%$. 
{The analysis with a moderate foreground removal model 
gives similar results as the optimistic case, 
and the difference in fractional errors of all four parameters is less than $2\%$.}

\section{Summary and future directions}
\label{sec:Discussion}
{\it Summary:} In this work, we studied the potential of upcoming redshifted 21~cm observations
from the SKA-Low to probe the primordial features predicted in inflation models. 
In particular, a class of inflation models based on particle physics
predict bump-like feature(s) on the primordial power spectrum 
associated with a burst(s) of particle productions during inflation
\cite{Chung:1999ve,Barnaby:2009mc,Barnaby:2009dd,Pearce:2017bdc}.
It was recently shown that such bursts of particle productions 
naturally occur in inflation models based on higher-dimensional gauge theories
\cite{Furuuchi:2015foh,Furuuchi:2020ery,Furuuchi:2020klq}.
CMB data from \textit{Planck} 2018
{gives} upper bounds on the amplitudes of bump-like features 
in the co-moving wave-number range
$0.0002 \lesssim k({\rm Mpc}^{-1}) \lesssim 0.15$ \cite{Naik:2022mxn}. 
In this work, we have investigated the possible constraints on
the parameters of bump-like features
{using mock data of SKA-Low}
in the co-moving wave-number range
$0.1 \lesssim k({\rm Mpc}^{-1}) \lesssim 1.0$. 
We focus on the  scenario of a single burst of particle production
that results in a single bump-like feature on the primordial 
power spectrum {over this range of co-moving wave-numbers,} 
parameterized by amplitude $A_{\rm I}$ and 
location $k_b({\rm Mpc}^{-1})$ via eq.~\eqref{eq:single_bump}. 
The mock 21~cm power spectra were created for the SKA-Low {antenna} configurations,
and Bayesian analysis with MCMC sampling was performed
to obtain posterior probability distribution on the parameters of interest. 

%
%
We first studied a scenario in which we varied 
only those parameters relevant to the primordial features, 
i.e., $A_{\rm I}$ and $k_b$, and fixed all the astrophysical parameters to 
their fiducial values. 
We used both optimistic and moderate foreground removal models 
to estimate the thermal noise from SKA-Low and the sample variance.
In addition to that, a modelling uncertainty of 20\% was added
to the noise power spectrum.
Our results are summarized case by case as follows (see figure~\ref{fig:all_results}):
\begin{itemize}
    \item Case I, input $(A_{\rm I}=10^{-9} , k_b=0.1({\rm Mpc}^{-1}))$:
single bump parameters $A_{\rm I}$ and $k_b$  were recovered within $68\%$ CL 
{of their input values}.  When the foreground removal model was changed from 
optimistic to moderate case, the change in fractional uncertainties on the parameters was 
less than $0.1\%$. 
    \item Case II, input $(A_{\rm I}=5\times10^{-10} , k_b=0.07({\rm Mpc}^{-1}))$:
   both  the parameters were recovered within $68\%$ CL {of their input values}.
   Compared with the optimistic case, the moderate foreground removal model
   increases the fractional errors on $\log A_{\rm I}$ and $k_b$ by 
   $0.08\%$ and $0.4\%$, respectively.
    \item Case III, {input} $(A_{\rm I}=5\times10^{-11} , k_b=0.07({\rm Mpc}^{-1}))$:
    both the parameters were recovered 
    within $68\%$ CL but with larger uncertainties.
    Compared to the previous case of a bump with an order of magnitude larger amplitude 
    at the same location, the fractional errors on $\log A_{\rm I}$
    and $k_b$ increased by $4\%$ and $51\%$, respectively. 
    {Moreover, the recovered value of $A_{\rm I}$ shows mild bias towards the lower side, while that of $k_b$ shows a relatively stronger bias towards the higher side.}
    \item Case IV,  {input} $(A_{\rm I}=3\times10^{-11} , k_b=0.04({\rm Mpc}^{-1}))$:
    a single bump model with the chosen parameters was allowed in the
    CMB data by {\it Planck} 2018 \cite{Naik:2022mxn}. The results show that
    the location of the feature with such a small amplitude could be difficult to constrain 
    from SKA-Low mock data. However, $\log A_{\rm I}$ and $k_b$ have recovered 
    within $68\%$ and $95\%$ CL, respectively.  
    {The recovered value of $k_b$ shows a stronger bias towards the higher side compared to Case III. This trend indicates that the location of the bump becomes harder to constrain as the input value of $A_{\rm I}$ is decreased. We note that the bias is observed when we compare the input values with the recovered {\em median} values in the cases when the shape of the one-dimensional PDF deviates strongly from the Gaussian form. We find that the maxima of the two-dimensional posterior PDFs are biased but still within 68\% CL from the input parameter values.}
    \end{itemize}

In the next scenario, we attempted to
simultaneously constrain 
astrophysical parameters along with the single bump model parameters. 
In principle, modelling the formation of the first stars, the heating of IGM, 
and the reionization of the universe involves many underlying astrophysical parameters. 
More unknown parameters make the parameter sampling computationally very expensive,
{particularly when the initial conditions need to be re-generated 
every time a new cosmological parameter is proposed.}
{To avoid detailed modelling of the EoR and 
to perform the analysis in a reasonable computational time, 
we considered a simple model of EoR, 
in which we varied} two of the astrophysical parameters: ionizing efficiency $\zeta$ and
minimum virial temperature of star-forming halos $T_{\rm vir}$ (K).
We added a modelling uncertainty of $20\%$ to the noise power spectra in our analysis. 
{The input parameters for creating mock power spectra are: 
$\zeta=30, \log T_{\rm vir}=4.69, A_{\rm I}=5\times10^{-10} {\rm and~ } k_b=0.07({\rm Mpc}^{-1})$.}
We found that the parameters $\zeta$ and $k_b$ have recovered
within $68\%$ CL, 
whereas $\log T_{\rm vir}$ and $\log A_{\rm I}$
have recovered within $95\%$ CL.
Compared with Case II (optimistic), 
we found that keeping $\zeta$  and $\log T_{\rm vir}$ as free parameters 
increases the fractional errors on $\log A_{\rm I}$ by $4\%$ and 
$k_b$ by $56\%$. 
\vspace{1mm}

\noindent\textit{Future directions:} We note that  particle production during inflation 
also leaves a bump-like feature on the bispectrum \cite{Barnaby:2010sq,Pearce:2017bdc}. 
Therefore, the bispectrum information from the
future redshifted 21~cm signal may further constrain the primordial features
discussed in this work.
{In the inflation models investigated here, 
multiple bursts of particle production can also occur during
inflation, leading to a {\em multi-bump} or a series of bump-like 
features on the primordial power spectrum. 
Such models contain a parameter that quantifies the spacing between the bumps 
in addition to the amplitude and location of the first bump.}  
{As  structure formation proceeds, 
it can be non-trivial to trace the 
signatures of different multi-bump models
only using the information from 
the two-point correlation functions. 
Therefore, joint analysis {with} higher-order correlators
 may be better suited to constrain the multi-bump models. 
We leave this case for future work. }

The results presented in this work can be improved further
if the estimated uncertainties are lowered, 
e.g. by increasing the observational hours or
increasing the effective area of the telescope\footnote{{E.g., the second phase of the SKA will have a larger collecting area than SKA1 \cite{Weltman:2018zrl}}.}. 
One can further reduce the modelling uncertainty by 
calibrating to more detailed radiative transfer simulations. 
There are other ways to characterize the redshifted 21~cm signal 
discussed in the literature (see \cite{Mesinger:2019cosm.book}
for an overview):
apart from the mean of $\delta T_b(\vec{x})$ given by 
the global 21~cm signal, there are one-point 
{geometrical and topological analysis of the 21~cm signal 
(see e.~g.~\cite{Kapahtia:2017qrg,Kapahtia:2021eok}).}
Therefore, a joint analysis using various statistical measurements may be necessary 
to simultaneously constrain all the EoR and cosmological parameters, 
including that of primordial features.
{Such an analysis will be carried out in the future as an extension of this work.} 
{It is also expected that the combined analysis of redshifted 21~cm observations
with }
futuristic observations from {other complementary probes, such as} CMB polarization, 
large-scale structure surveys and gravitational waves,
would provide the prospect of further constraining the inflation models. 

\section*{Acknowledgements}
\label{sec:ack}
We thank the anonymous referee for the helpful comments/suggestions. 
SSN thanks the Indian Institute of Astrophysics, Bangalore,
for the hospitality and support provided during an academic visit.
SSN also thanks Debbijoy Bhattacharya for valuable inputs and stimulating discussions.
{The computational work in this paper was carried out using  {\it Shakti,} 
a high-performance computing cluster 
at the Manipal Centre for Natural Sciences, Centre of Excellence,
Manipal Academy of Higher Education. }

This work was supported in part by 
Dr~T.M.A.~Pai Ph.~D. scholarship program of 
Manipal Academy of Higher Education
and the Science and Engineering Research Board,
Department of Science and Technology, Government of India,
under the project file number EMR/2015/002471.

\appendix
\section{Breaking potential degeneracy between the single bump model parameters and the EoR parameters}
\label{app:degeneracy}
%
\begin{figure}[!h]
	\centering
	\includegraphics[width=0.8\textwidth,height=6cm]{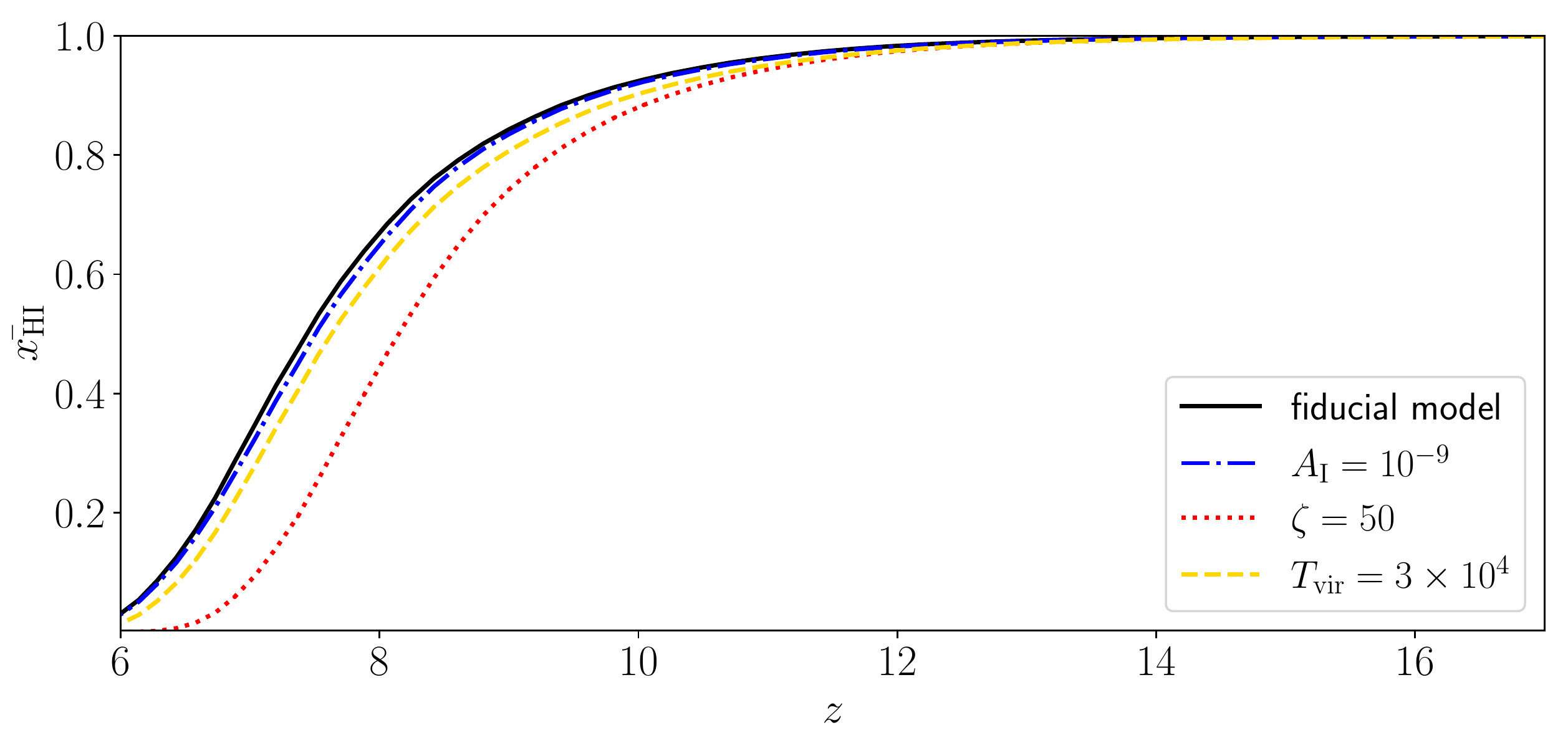}
	\caption{The average neutral fraction of hydrogen versus redshift $z$ for the models under consideration.}
	\label{fig:xHI_vs_z}
\end{figure}
In this section, we discuss the possible degeneracy
 between the amplitude of the bump-like primordial features
$A_{\rm I}$ and 
the parameters relevant for EoR ($\zeta$, $T_{\rm vir}$).
As discussed in section~\ref{subsec:result_EoR}, 
increasing $\zeta$ has similar effects on structure formation 
as decreasing $T_{\rm vir}$ and 
as increasing power on the primordial power spectrum
by, e.g., adding a feature with amplitude $A_{\rm I}$.
In order to examine their correlation further, we consider how the following models differ from the fiducial one: 
\begin{itemize}
    \item {\bf Model {\it a}:} increasing $\zeta$ from 30 (fiducial) to 50. 
    \item {\bf Model {\it b}:} decreasing $T_{\rm vir}$ from $5\times 10^4$ (fiducial) 
    to $3\times 10^4$. 
    \item {\bf Model {\it c}:} adding a bump-like primordial feature of 
    amplitude $10^{-9}$ at $k_b=0.05$. 
\end{itemize}
The remaining parameters in the simulation are fixed to their benchmark 
values given in section~\ref{subsec:21cm}. 
The corresponding ionization history encoded in the mean neutral fraction of hydrogen 
$\bar{x_{\rm HI}}$ versus $z$ for each model is shown in figure~\ref{fig:xHI_vs_z}. 
We observe that introducing a bump feature quickens the ionization of the universe, 
similar to increasing $\zeta$ or decreasing $T_{\rm vir}$.  
The figure, therefore, highlights the degeneracy between the three parameters in the ionization history.

\renewcommand\thesubfigure{\roman{subfigure}}
\begin{figure}[!tbp]
    \centering
    \subfloat[21~cm power spectra]{{\includegraphics[height=10cm,width=16cm]{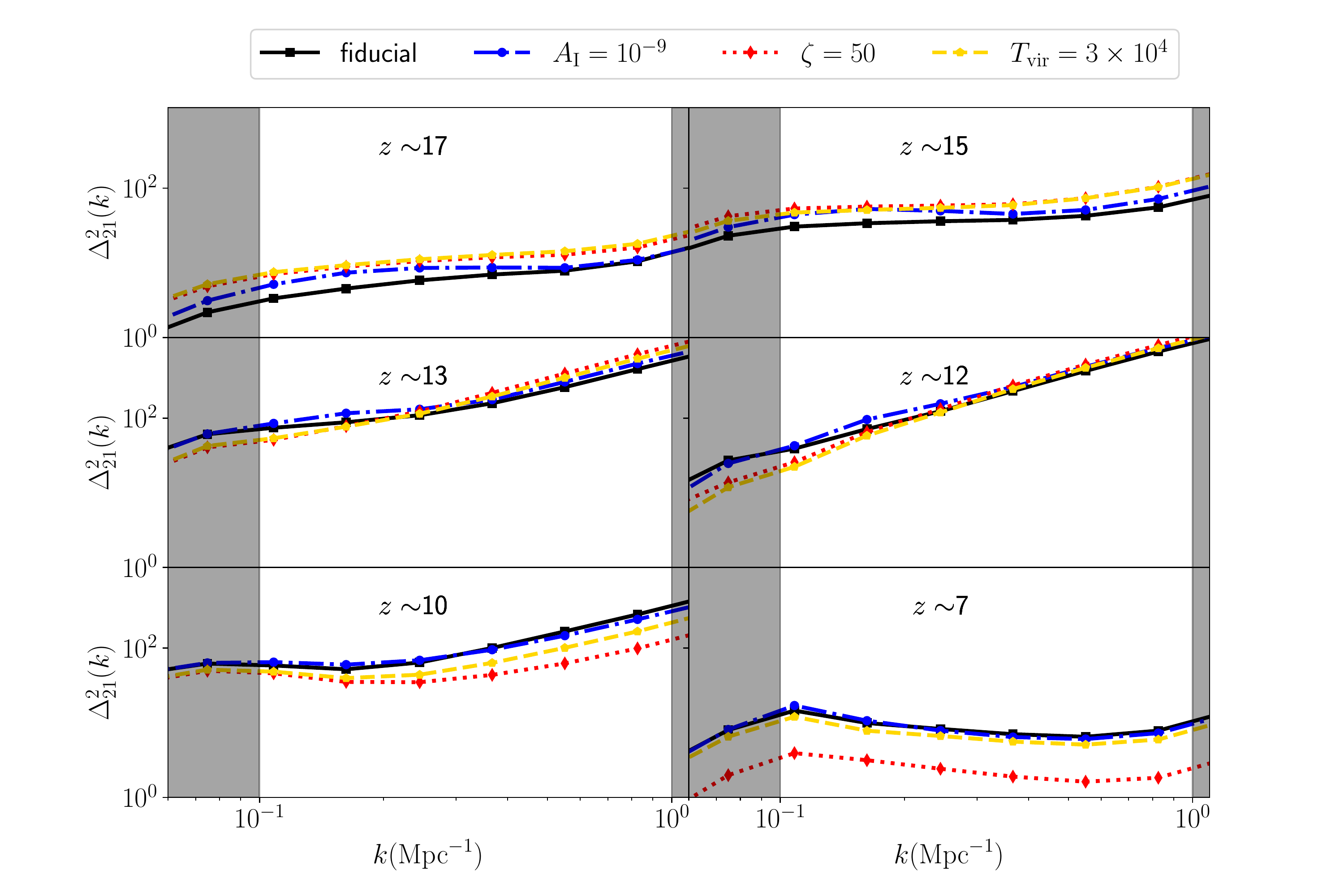}}}\\
    \subfloat[Fractional difference of the power spectra]{{\includegraphics[height=10cm,width=16cm]{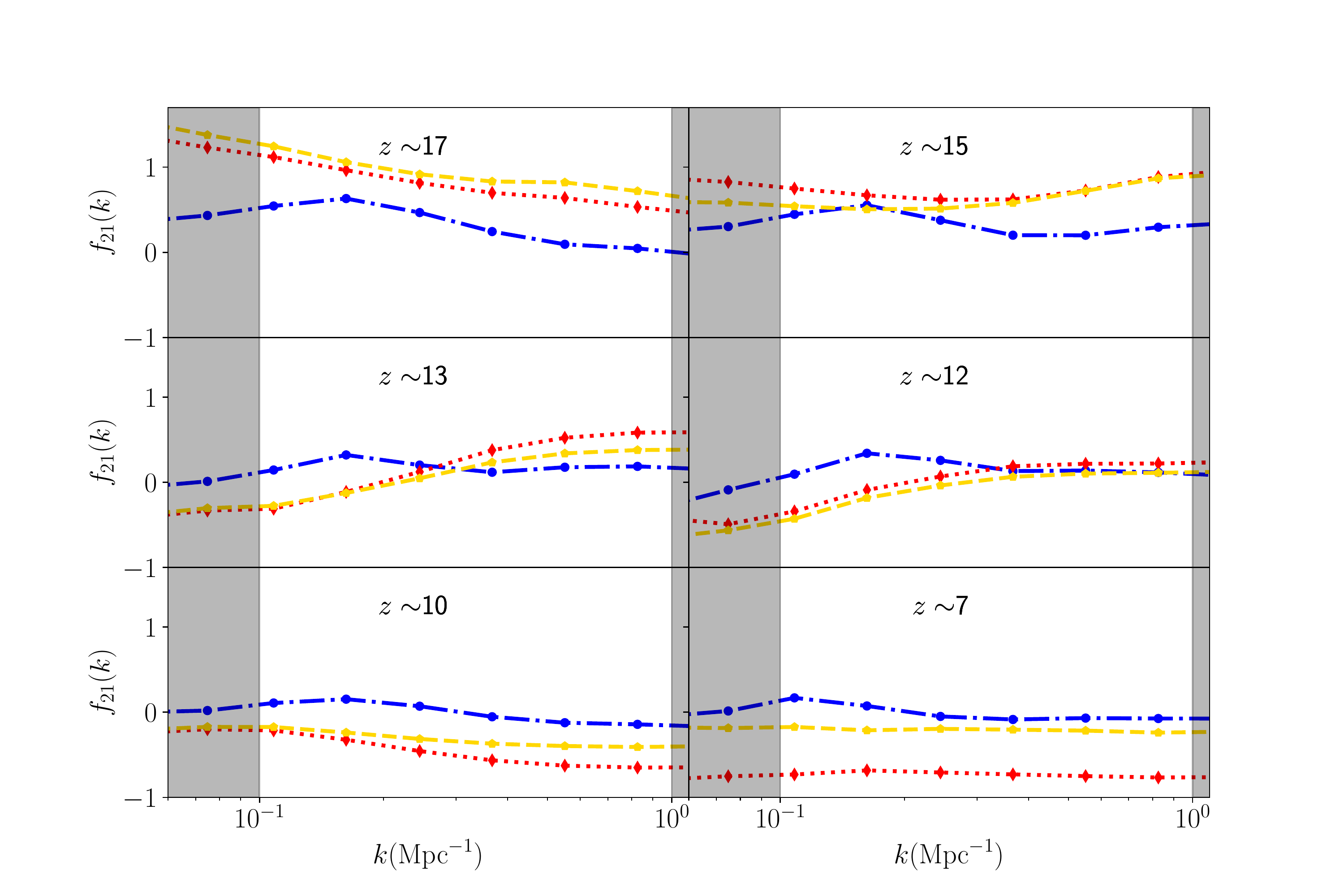}}}
    \caption{\small{The simulated 21~cm power spectra for the fiducial model (featureless, $\zeta=30$ and $T_{\rm vir}=5\times 10^4$)
    are shown by solid lines. 
    Models {\it a} (featureless, $\zeta=50$), {\it b} (featureless, $T_{\rm vir}=3\times 10^4$), and
    {\it c}  (single bump with $A_{\rm I}=10^{-9}$, $k_b=0.05$) are shown by dotted, dashed, and dot-dash lines, respectively. 
    The simulated volume in \texttt{21cmFAST} has a box length of 250 Mpc with $84^3$ grids. 
    The grey-shaded regions are excluded from our MCMC sampling.}}
    \label{fig:degeneracy}
\end{figure}

To distinguish the models {\it a}, {\it b} and {\it c}, we next discuss their 21 cm power spectra. 
To quantify how each model differs from the fiducial one, we define the fractional difference of the power spectra, $f_{21}$, as
\begin{equation}
f_{21}(k):= \frac{\delta\Delta^2_{21}(k)}{\Delta^{2,{\rm fid}}_{21}(k)} = \frac{\Delta^{2}_{21}(k) -\Delta^{2,{\rm fid}}_{21}(k)}{\Delta^{2,{\rm fid}}_{21}(k)}\,,
\end{equation}
where the suffix `fid' stands for the fiducial model. 
In figure~\ref{fig:degeneracy}, we plot $\Delta^2_{21}$ (top panels) 
and $f_{21}$ 
(lower panels) as functions of $k$ at a few redshifts for all the  models. 
We note the following points:
\begin{itemize}
	\item The introduction of the bump enhances the power towards the lower $k$ side (towards the location of the bump but not coincident), relative to the fiducial model, at all redshifts. This enhancement is distinct from the effects of varying $\zeta$ and $T_{\rm vir}$, which are seen to {(i)} enhance power at all $k$ at $z\sim 17$ and $z\sim 15$, {(ii)} enhance (suppress) at high (low) $k$ values at $z\sim 13$ and $z\sim 12$, and {(iii)} suppress power at all $k$ for $z\sim 10$ and $z\sim 7$.

     \item 
     {At relatively low redshifts  ($z\sim 10$ and $z\sim 7$), the fractional difference drops down, and  the bump model becomes hard to distinguish from the fiducial model, which demonstrates that the effect of the primordial bump-like feature is  washed out towards the lower redshifts where the astrophysical details become very important. This is an important point to note when targeting to constrain primordial physics using the 21 cm power spectrum.  In comparison, the different EoR models are quite distinct from the fiducial model.}

\end{itemize}

In conclusion, we find that the effects on 
the 21cm power spectra due to adding 
bump-like features become almost insignificant
at the lower redshifts ($z<10$), 
making this redshift regime an ideal window 
for constraining the EoR parameters. 
On the higher redshifts ($z>10$), 
where the  bump-like features have a non-negligible effect, 
the distinct profiles of the 21cm power spectrum 
due to the change in EoR parameters and 
the addition of primordial feature make this redshift regime
suitable to extract the signatures of primordial features.
Thus, with the knowledge of EoR parameters 
obtained from multiple 21cm observations 
in the lower redshift ($z<10$) and 
the distinct profiles of 21cm power spectra 
at higher redshifts, 
it is possible to break the degeneracy 
between the parameters of EoR and primordial features. 
This is an encouraging finding from the point of view of using 21 cm observations to constrain primordial physics, such as particle production during inflation. However, the caveat is that observed data towards higher redshifts, where instrumental noise tends to increase,  are required.


\def\apj{ApJ}%
\def\mnras{MNRAS}%
\def\aap{A\&A}%
\def\apjl{ApJ}
\def\aj{AJ}
\def\physrep{PhR}
\def\apjs{ApJS}
\def\jcap{JCAP}
\def\pasa{PASA}
\def\pasj{PASJ}
\def\nat{Natur}
\def\apss{Ap\&SS}
\def\araa{ARA\&A}
\def\aaps{A\&AS}
\def\ssr{Space Sci. Rev.}
\def\pasp{PASP}
\def\na{New A}

\bibliography{ref_21cm.bib}
\bibliographystyle{JHEP}

\end{document}